\documentclass[10pt,newlfont,pra,aps,superscriptaddress,twocolumn]{revtex4}
\usepackage[pdftex]{graphicx}
\usepackage{here}
\usepackage{amsmath}
\usepackage{atbeginend}
\usepackage{exscale}
\def\hs{\hspace{0.5cm}}
\def\xsand{$^4$He-$^3$He}
\def\xhe3{$^3$He}
\def\zhe4{$^4$He}
\def\he3he3{${}^3$He-${}^3$He}
\def\che4he4{${}^4$He-${}^4$He}
\def\ea{{\it et al.} }
\def\ira{{\it Ira}}
\def\masha{{\it Masha}}
\def\sandw{\xhe3-\zhe4}
\def\ft{\hspace{1mm}}
\pdfoutput=1
\begin{document}
\title{A Comprehensive Study of the $^3$He-He II Sandwich
System Using Monte Carlo Techniques}

\author{Amer Al-Oqali}
\affiliation{Department of Physics, Faculty of Science, The University of Jordan, 
Amman, JORDAN}
\author{Asaad R. Sakhel}
\affiliation{Faculty of Engineering Technology, Al-Balqa Applied University 
Amman 11134, JORDAN}
\author{Humam B. Ghassib$^1$}
\begin{abstract}
\hs We present a numerical investigation of the thermal and structural
properties of the \xsand\ sandwich system adsorbed on a graphite substrate
using the Worm Algorithm Quantum Monte Carlo (WAQMC) method 
\cite{Boninsegni:06a}. For this purpose, we modified a previously written
WAQMC code originally adapted for \zhe4\ on graphite, by including the 
second \xhe3-component. In order to describe the fermions, a 
temperature-dependent statistical potential was used which proved very 
effective. To the best of our knowledge, the statistical potential has 
not been used before in Quantum Monte Carlo techniques for describing
fermions. In an unprecedented task, the WAQMC calculations were conducted 
in the milli-Kelvin temperature regime. However, because of the heavy 
computations involved, only 30, 40, and 50 mK were considered for the 
time being. The pair correlations, Matsubara Green's function, structure 
factor, and density profiles were explored at theses temperatures. (Note: 
this paper is just a preliminary version and will be replaced by an 
updated version.)
\end{abstract}
\date{\today}
\maketitle

\section{Introduction}

\hs There have been only a few investigations on the \xsand\ sandwich 
system in the last 25 years \cite{Ghassib:1994,McQueeney:1984}, 
most of the studies having concentrated on \xsand\ films 
\cite{Akimoto:2003,Akimoto:2006,Ziouzia:2003,Nash:2003,Ghassib:1984} 
and superfluid \zhe4\ films \cite{Pierce:1999,Pierce:2001}. These 
investigations aimed at calculating the Fermi liquid parameters, 
the speed of third sound in He II, the specific heat capacity, and 
the Kosterlitz-Thouless (KT) transition. \xsand\ mixtures and 
films \cite{Krotscheck:2001,Ghassib:1984a} are considered important 
physical systems for several reasons: 1) their use in cooling to 
the milli-Kelvin regime; 2) the central role as theoretical labs
for the study of a number of methods in many-body physics; and 3) 
the importance of the sandwich system specifically in its role where 
dimensionality effects arise. One can thus see the importance of 
this study, particularly since it will be conducted using Quantum 
Monte Carlo techniques. Previous work on \xsand\ mixtures and films 
is abundant. Experimentally, the torsional oscillator was used 
to study the superfluid \xsand\ sandwich system \cite{McQueeney:1988} 
and it was found that the critical temperature for the Kosterlitz-Thouless 
(KT) transition decreases as the number of \xhe3\ atoms is increased. 
Measurements on third sound in \xsand\ films have also been 
conducted. It was found that by increasing the concentration of \xhe3\ 
in  \zhe4, the speed of third sound decreases and a complete phase 
separation occurs at $T\le 0.5$ K. As a result, this system resembles 
the \xsand\ sandwich system. Ghassib and Waqqad \cite{Ghassib:1990} 
reconsidered Bose-Einstein condensation in an ideal, quasi two-dimensional 
Bose gas and explored crossover effects from two- to three-dimensional 
systems. Further, Ghassib and Chatterjee \cite{Ghassib:1984a} examined 
the effects of \zhe4\ impurities on some low-temperature properties of 
normal liquid \xhe3. It was argued that no \xsand\ mixtures can 
possibly exist at very low temperature ($T\le 100$ mK), where a total 
phase separation occurs.

\hs From another point of view, the possibility for dimer and trimer 
formation in \xsand\ films was explored. Ghassib \cite{Ghassib:1984}
predicted that dimers form initially in \xhe3; afterwards $-$at much 
lower temperatures$-$ a KT transition could occur for boson composites.
\xsand\ mixtures in two dimensions have also been considered. For 
example Krotscheck \ea\ \cite{Krotscheck:2001} showed than an effective 
interaction between pairs of \xhe3\ atoms inside a host \zhe4\ liquid 
was sufficient to cause loosely-bound dimers.

\hs Investigations of \zhe4\ on a graphite substrate have also been 
conducted. For example, Corboz \ea\ \cite{Corboz:2008} investigated 
the low-temperature phase diagram of the first and second layer of \zhe4\ 
adsorbed on graphite, using the worm algorithm. Pierce and Manousakis 
\cite{Pierce:1999,Pierce:2001} presented a path-integral Monte Carlo 
(PIMC) method for simulating helium films on the graphite surface, and 
investigated helium layers adsorbed on the substrate. In addition, 
diffusion Monte Carlo has also been used to study the first layer of 
\zhe4\ adsorbed on graphite \cite{Gordillo:2009}, and the ground-state 
properties of the homogeneous two-dimensional liquid \zhe4\ 
\cite{Giorgini:1996}. 

\hs It is obvious that these previous investigations are not 
enough; here here we provide a more comprehensive in depth 
microscopic study of this system. Our chief goal is to compute
some thermal and structural properties of the \xsand\ sandwich
system in the milli Kelvin temperature regime using the Worm
Algorithm Quantum Monte Carlo method \cite{Boninsegni:06a}. To
the best of our knowledge, this kind of system has not been
simulated before in such a low-temperature regime. Because of 
the heavy computational aspect of the present simulations, we 
were only able to obtain results for three temperatures: $T=30$, 
40, and 50 mK.

\hs The \xsand\ sandwich system proper consists of a \zhe4\ solid layer 
of $\sim 3.6\AA$ thickness adsorbed on the walls of a container, above 
which resides a \xsand\ mixture-layer of 7-11 $\AA$ thickness followed 
by a pure bulk liquid \xhe3\ layer. In this paper, we rejuvinate the 
investigations on this sandwich system which promises richness in physics. 
We chiefly investigate the thermal properties, such as the pressure, 
internal energy, entropy, and superfluid density. By using the superfluid 
density one can detect the role of \xhe3\ atoms in the depletion of the
superfluid in such a many-body system. In addition, other properties 
can be obtained such as the solubility of \xhe3\ into \zhe4, and the 
density profiles which show layer promotion upon increasing the number 
of \xhe3\ or \zhe4\ atoms. Further, the Matsubara Green's function 
$G(p,\tau)$ \cite{Mahan:1990} is computed by the numerical implementation
of the WAQMC code \cite{Boninsegni:06a} in order to check for excitations
and particle propagation in the sandwich system. 
Another key point is 
that we use a statistical potential \cite{Pathria:1996} in order to 
include real fermionic statistics into the calculations, thereby 
circumventing the fermion sign problem, which would otherwise arise 
if we allowed sign-changes corresponding to permutations of the fermions. 

\hs We thus consider $N$ \xhe3\ and \zhe4\ atoms with different numeric
ratios in a \xsand\ sandwich system on graphite. The interactions between
the \zhe4\ atoms and the \xsand\ pairs are described by the Aziz potential 
\cite{Aziz:1979}; whereas the ${}^3$He atoms interact by the fictitious 
statistical potential. The Worm-Algorithm Quantum Monte Carlo (WAQMC) 
method \cite{Boninsegni:06a} is used to simulate this system. For this 
purpose, we modified a previously written Worm-Algorithm code 
\cite{Nikolay:2009} specifically designed for \zhe4\ on graphite, by 
including a second component (\xhe3) into the code. The use of a statistical 
potential in the description of fermions in a Monte Carlo simulation is 
unprecedented, and we hope to be able to convince the reader of its 
effectiveness.

\hs We found chiefly that the statistical potential is very effective
in describing the \xhe3\ fermions in a \zhe4\ environment. The pair
correlation function reveals strong correlations between the three pairs
of \he3he3, \che4he4, and \xsand\ atoms, signalling the presence of 
different types of clusters. The Matsubara Green's function demonstrated
substantial activity in the system and a condensate fraction as well.
The integrated density profiles, taken in a plane perpendicular to the
substrate, revealed crystallization of the layers closest to the graphite
substrate; whereas disorder is prevalent in the layers father away from 
the substrate.

\hs The organization of the paper is as follows. In Sec.\ref{sec:method}
we describe the changes we made in the WAQMC code so as to include
the \xhe3\ component. For this purpose, we needed to recast some of the
information in Ref.\cite{Boninsegni:06a}, so that the reader can
understand our changes. In Sec.\ref{sec:resanddis} we present the results
of our calculations and discuss them. Finally, in Sec.\ref{sec:conclusions}
we present our conclusions.

\section{Method}\label{sec:method}

\hs In this section, we do not explain the WAQMC technique; we only 
outline our modifications to the code. The WAQMC method has been explained in 
detail by the inventors of the technique \cite{Boninsegni:06a}. This technique 
is relatively new and based on conventional path integral Monte Carlo (PIMC) 
described earlier by the excellent review of David Ceperley \cite{Ceperley:95}. 
The idea behind the WAQMC method was to make PIMC more efficient by introducing 
off-diagonal configurations (worms) into the system in addition to the existing 
diagonal ones. That is, in WAQMC one uses configurations containing both closed 
(diagonal) world-lines and {\it one} open (off-diagonal) world-line (worm). 
The diagonal configurations contribute to the partition function, hence referred 
to as the $Z-$sector, whereas the off-diagonal ones to the Matsubara Green's 
function, G, hence referred to as the $G-$sector \cite{Boninsegni:06a}. In PIMC 
as well as WAQMC, each particle is represented by a trajectory in space-time 
which closes upon itself in space after it has moved for a time $\beta$. Each 
position in space-time on this trajectory is represented by a bead, and each 
pair of consecutive beads is separated by a time slice $\tau$. If there are $M$ 
time slices, then $\beta=M\tau$, where $M$ is the number of time slices along a 
certain trajectory, the path of the particle in space-time. The particle is thus 
described by a ring-polymer, an entirely new picture \cite{Ceperley:95}.

\subsection{Interactions}

\hs For the \che4he4\ and \xsand\ interactions, the standard interatomic
Aziz potential \cite{Aziz:1979} was used. For the \he3he3\ interactions
we invoked a fermionic statistical potential \cite{Pathria:1996} given by

\begin{equation}
v_s(r)\,=\,-k_B T \ln[1-\exp(-2\pi r^2/\lambda^2)], 
\label{eq:statisticalp}
\end{equation}

where $\lambda=\hbar^2/(2m)$, $k_B$ being Boltzmann's constant, $r$ the
distance between a pair of \xhe3\ atoms, and $T$ the temperature. The
idea behind the statistical potential is to simulate real fermions;
thereby circumventing the sign problem, as mentioned previously.

\subsection{Worm updates}\label{sec:worm-updates}

\hs In what follows, we describe the changes that we implemented in the 
WAQMC code in order to include the second \xhe3\ component. For this purpose, 
we recast some of the information in Ref.\cite{Boninsegni:06a} in order to shed 
enough light on the changes. All the worm-update equations and concepts
used in this paper were given earlier in Ref.\cite{Boninsegni:06a},
except for the indicated changes made to accommodate the \xhe3\
component.

\hs The worm updates are accepted or rejected according to certain, 
carefully defined probabilities. In essence, only {\it one} worm is 
allowed and added to the diagonal configurations. This worm, when 
inserted, has a starting bead named for historical reasons \masha\ 
($\cal M$), and an ending bead named \ira\ ($\cal I$). \ira\ always 
advances \masha\ in time. \ira\ or \masha, that is the end-beads of a 
worm, can be moved forward or backward in time. They can be reconnected 
to diagonal trajectories after these trajectories are cut open, and 
they can also close an off-diagonal trajectory by glueing a worm to 
the opening. Further, a worm can be erased and then reintroduced. Beads 
across two different trajectories can be linked together by diagrammatic 
links, leading to bonds between them. 

\hs The WAQMC code was originally written \cite{Boninsegni:06a} for one 
component only, namely \zhe4, on graphite. To include the \xhe3\ 
component, a logical array $who_{-}are{}_{-}you(bead)$ was introduced 
which would return a {\it true} value for a chosen bead if it was \zhe4, 
and {\it false} if it was \xhe3\ in order to label the particles and to 
distinguish between them.

\begin{figure}[t!]
\includegraphics*[width=8.5cm,viewport=101 561 519 783]{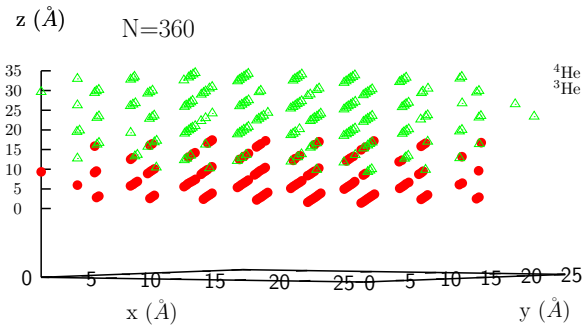}
\caption{\footnotesize\baselineskip 0.2cm Initialization of a
\xsand\ sandwich system of $N=360$ atoms on a graphite substrate.
The \zhe4\ atoms (red circles) are adsorbed on a graphite surface
constituting a layer of $\sim 3.6\AA$ thickness. The \xhe3\ atoms 
(green triangles) form a bulk layer of about $\sim 7\AA$ thickness.
Sandwiched in between these two is a \xsand\ mixture-layer of 
$\sim 15\AA$ thickness. The ratio of \xhe3\ and \zhe4\ atoms in 
the latter is chosen randomly.}\label{fig:initialization}
\end{figure}

\subsubsection{Initialization}

\hs The \xsand\ sandwich system is initialized using straight 
world-lines each of length $\beta$ as shown in Fig.1, for a system 
of, e.g., $N_3=222$ \xhe3\ atoms and $N_4=138$ \zhe4\ atoms. A logical 
bead list $who_{-}are{}_{-}you(bead)$ is initialized as the sandwich 
system is built up into layers on graphite. The first layer adsorbed 
on the graphite surface consists of \zhe4\ atoms only constituting 
about 25$\%$ of the total number of atoms $N$ ($0.25\beta/\epsilon$ 
beads), the second consists of a \xsand\ mixture constituting 25$\%$ 
of $N$, whereas the third layer consists only of \xhe3\ atoms constiuting 
the rest of $N$. Here, $\epsilon$ is the ``time step" in the Worm 
Algorithm technique. The type of the atoms in the mixture-layer is
randomly assigned to simulate a realistically mixed layer.

\begin{figure}[t!]
\includegraphics*[width=7cm,viewport=90 347 297 712]{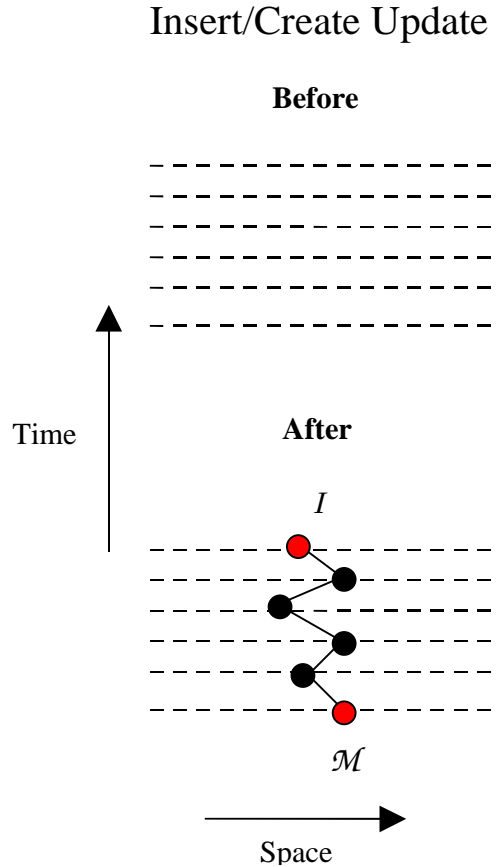}
\caption{\baselineskip 0.2cm Worm-Algorithm Insert update representation 
in space-time coordinates. \ira\ ($\cal I$) and \masha\ ($\cal M$) are show 
as red solid circles.}\label{fig:insert}
\end{figure}

\subsubsection{Insert}\label{sec:insert}

\hs A worm, either fermionic or bosonic, is created as shown in 
Fig.\ref{fig:insert}, where the beginning of the worm is 
\masha\ ($\cal M$) and the end \ira\ ($\cal I$). The figure is 
a presentation of the open (off-diagonal) trajectory in space-time. 
The type is assigned randomly using a certain probablity: If a random 
number, $\xi< 0.5$ say, the mass used in the updates will be that of \zhe4; 
if $\xi\ge 0.5$, the mass is that of \xhe3. Accordingly, we use 
in FORTRAN 90

\begin{center}
\begin{eqnarray}
&&\xi=rndm()\nonumber\\
&&\mathbf{IF}\hspace{0.cm}(\xi\,\,\mathbf{.lt.}\,\,0.5)\hspace{0.cm}
\mathbf{THEN}\nonumber\\
&&mp\,=\,m4\nonumber\\
&&\mathbf{ELSE\hspace{0.cm}IF}\hspace{0.cm}(\xi\,\,\mathbf{.ge.}\,\,0.5)
\hspace{0.cm} \mathbf{THEN}\nonumber\\
&&mp\,=\,m3\nonumber\\
&&\mathbf{ENDIF}
\end{eqnarray}
\end{center}

where $mp$ is the mass variable in the program, and $m3$ and $m4$ 
are the masses of \xhe3\ and \zhe4. In the upcoming types of worm 
updates, the beads, newly created or removed, are assigned the value 
{\bf .TRUE.} or {\bf .FALSE.}, respectively, according to the choice 
of the mass in the INSERT update above. Thus except for the CUT 
update (see Sec.\ref{sec:cut} below), the types of beads and the 
associated mass used is the same as that chosen initially in the 
INSERT update.

\hs The acceptance probability for this INSERT update is 
(as in Ref.\cite{Boninsegni:06a})

\begin{equation}
P_{in}\,=\,min\left\{1,2CVP\overline{M}\,e^{\Delta U\,+\,\mu 
M \epsilon}\right\},
\end{equation} 

where $\Delta U$ is the change in the configurational potential energy 
of the beads due to the insertion of the worm, $\mu$ the chemical potential, 
and $\epsilon$ is the time step. Here, $C$ is a constant, $V$ the volume of 
the system, $M$ the length of the worm proposed which is selected randomly 
within an interval $[1,\overline{M}]$, and $P=\beta/\epsilon$ is the number 
of time slices along the path of ``length" $\beta$. In the WAQMC code, 
$P_{in}$ is programmed as follows:

\begin{equation}
P_{in}\,=\,w_{ST}\cdot w_t \cdot V\,\frac{\beta}{2}\,\frac{\overline{M}}{2}\,
\frac{p_{re}}{p_{in}} \,e^{\mu\epsilon M +\Delta U}, \label{eq:Pin} 
\end{equation}

where, $w_{ST}$ controls the worm statistics, $p_{re}$ and $p_{in}$ are 
fixed attempt probabilities for removing and inserting a worm, respectively, 
and $w_t$ is a weight determined from the total number of beads before and 
after an update. We multiplied Eqs.(\ref{eq:Pin}) and (\ref{eq:Pcut}) below 
by $1/p_f$ or $1/p_b$ for a fermion or boson worm, respectively, where 
$p_f$ is the attempt probability for getting a fermion and $p_b=1-p_f$ the 
attempt probability for getting a boson.

\subsubsection{Remove}

\begin{figure}[t!]
\includegraphics*[width=7cm,viewport=94 365 298 713]{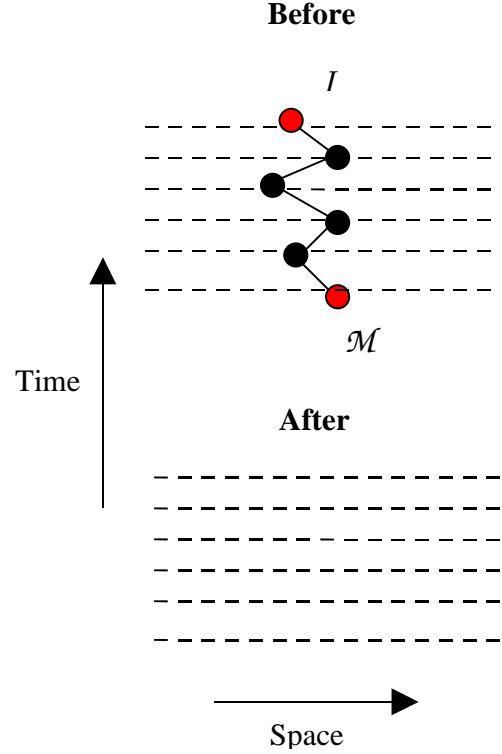}
\caption{\baselineskip 0.5cm As in Fig.\ref{fig:insert}; but for 
the removal of a worm.}
\label{fig:remove}
\end{figure}

\hs A worm, either fermion or boson, is removed (annihilated) as shown in 
Fig.\ref{fig:remove}. The type of worm to be removed depends on the 
mass $mp$ chosen in the INSERT update above. That is, if $mp=m3$, 
then a fermion worm is removed, otherwise if $mp=m4$ a boson worm. 
The probability for this update is

\begin{equation}
P_{rm}\,=\,min\left\{1, \frac{e^{\Delta U-\mu \overline{M} 
\epsilon}}{2CVP\overline{M}}\right\},
\end{equation}

and in the WAQMC program it is coded

\begin{equation}
P_{rm}\,=\,\frac{4\,w_t\,e^{-\mu \epsilon \overline{M}+\Delta U}
\,p_{in}}{w_{ST}\cdot V\cdot \beta \overline{M} p_{re}}.\label{eq:Pcut}
\end{equation}

As a preventive measure during the process of removing the beads,
if at any time a bead to be removed has a different type than the 
worm beads on which the update is performed, the program 
terminates. But this is just in case and is not supposed to
happen.

\subsubsection{Move Forward Masha}

\begin{figure}[t!]
\includegraphics*[width=7cm,viewport=94 224 317 678]{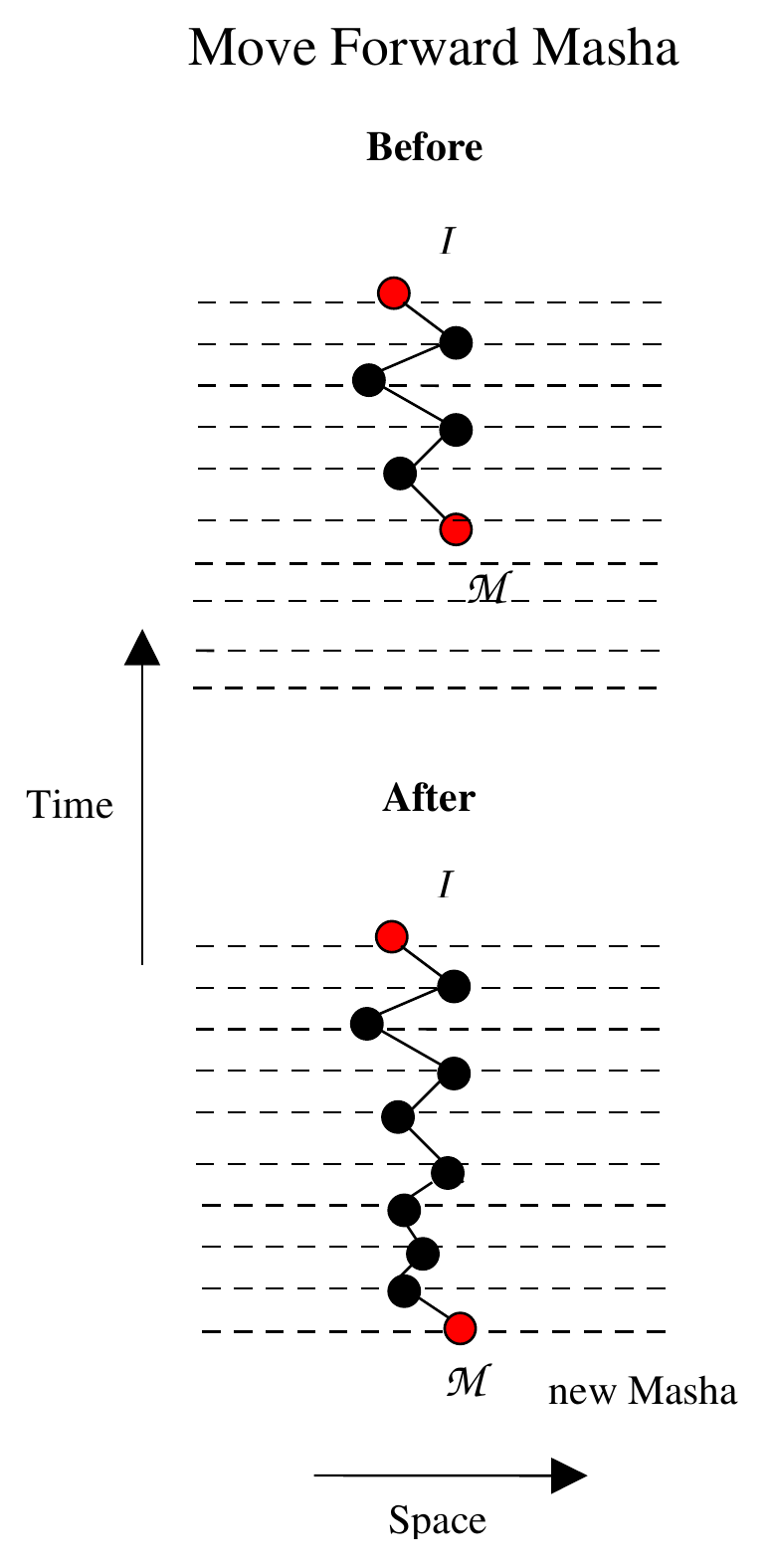}
\caption{\baselineskip 0.5cm Worm Algorithm update for moving \masha\
forward}\label{fig:MoveForwardMasha}
\end{figure}

\hs In this update, the beginning of the worm (timewise speaking slice 
number 0) is propagated backwards in time as shown in 
Fig.\ref{fig:MoveForwardMasha}. That is to say, a chain of new beads
is attached to the old \masha\ backwards in time ending then with
a new {\it Masha}. The old \masha\ is then relabelled as an ordinary
bead. In the event that a newly generated bead has a different type 
than \masha\, the program terminates according to the code:

\begin{center}
\begin{eqnarray}
&&\mathbf{IF}\hspace{0.0cm}( who{}_-are{}_-you(bead)
\hspace{0.0cm}\mathbf{.ne.}
\hspace{0.0cm} who{}_-are{}_-you({\cal M}))\hspace{0.0cm} 
\mathbf{STOP} \nonumber\\ 
\label{for90:stop.bead.M}
\end{eqnarray}
\end{center}

The type of the worm is pre-determined in the INSERT update. 
The probability for this update advancing \masha\ forward 
is

\begin{equation}
P_{ad, \cal M}\,=\,min\left\{1, e^{-\Delta U+\mu M \epsilon}\right\},
\end{equation}

and in the program it is coded

\begin{equation}
P_{ad,\,\cal M}\,=\,e^{\mu\epsilon\overline{M}-\Delta U}\cdot w_t 
\cdot w_{lc, \cal M}.
\end{equation}

Here $w_{lc, \cal M}$ is the worm-link correction of the links to 
\masha\ ($\cal M$):

\begin{equation}
w_{lc_{\cal M}}\,=\,\prod_{i=1}^{N_\ell}\,\left(\frac{\displaystyle 
e^{-\epsilon V(|\mathbf{r}_{\cal M}-\mathbf{r}_i|)}-1}
{e^{-\epsilon V(|\mathbf{r}_{\cal M}-\mathbf{r}_i|)/2}-1}\right),
\end{equation}

where $r_{\cal M}$ is the position of \masha, $\mathbf{r}_i$ the 
position of the bead $i$ linked to \masha, $N_\ell$ is the number of 
links to \masha, and $V(r)$ is the pair interaction potential. 
Here, it doesn't matter what type of bead one links to since
nothing prevents the formation of bonds between fermions and
bosons.

\begin{figure}[t!]
\includegraphics*[width=7cm,viewport=102 276 300 716]
{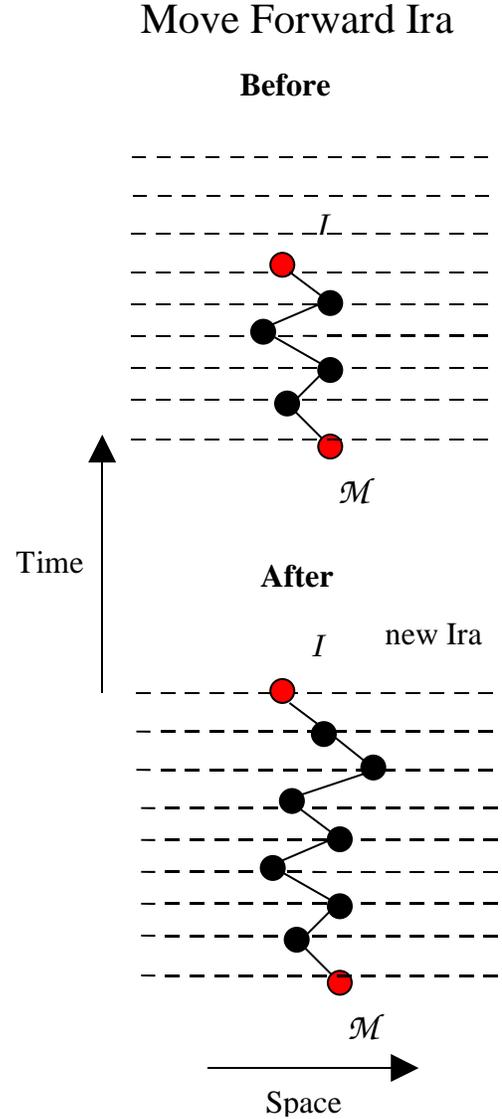}
\caption{\baselineskip 0.5cm Worm Algorithm update for moving 
\ira\ forward.}\label{fig:MoveForwardIra}
\end{figure}

\subsubsection{Move Forward Ira}

\hs In this update, the end of the worm (timewise speaking last slice 
on worm) is propagated forward in time as in Fig.\ref{fig:MoveForwardIra}. 
Again, the type of worm is pre-determined in the INSERT update, and any 
newly created beads must have the same type as that of the worm to be 
updated. If it happens that a bead has a different type than the worm, 
the program terminates according to

\begin{center}
\begin{eqnarray}
&&\mathbf{IF}\hspace{0.0cm}( who{}_-are{}_-you(bead)\hspace{0.0cm}
\mathbf{.ne.}\hspace{0.0cm} who{}_-are{}_-you({\cal I}))\hspace{0.0cm} 
\mathbf{STOP}. \nonumber\\ \label{for90:stop.bead.I}
\end{eqnarray}
\end{center}

The probability for this update is

\begin{equation}
P_{ad,\cal I}\,=\,min\left\{1, e^{-\Delta U+\mu M\epsilon}\right\},
\end{equation}

and is coded

\begin{equation}
P_{ad,\cal I}\,=\,e^{\mu\epsilon\overline{M}-\Delta U}\cdot w_t 
\cdot w_{lc_{\cal I}}.
\end{equation}

Here $w_{lc_{\cal I}}$ is the worm link correction of all the links 
to \ira\:

\begin{equation}
w_{lc_{\cal I}}\,=\,\prod_{i=1}^{N_\ell}\,\left(\frac{\displaystyle 
e^{-\epsilon V(|\mathbf{r}_{\cal I}-\mathbf{r}_i|)}-1}
{e^{-\epsilon V(|\mathbf{r}_{\cal I}-\mathbf{r}_i|)/2}-1}\right),
\end{equation}

where $\mathbf{r}_{\cal I}$ is the position of $\cal I$ and $N_\ell$ 
is the number of links to $\cal I$.

\subsubsection{Move Backward Masha}

\begin{figure}[t!]
\includegraphics*[width=7cm,viewport=94 242 327 674]
{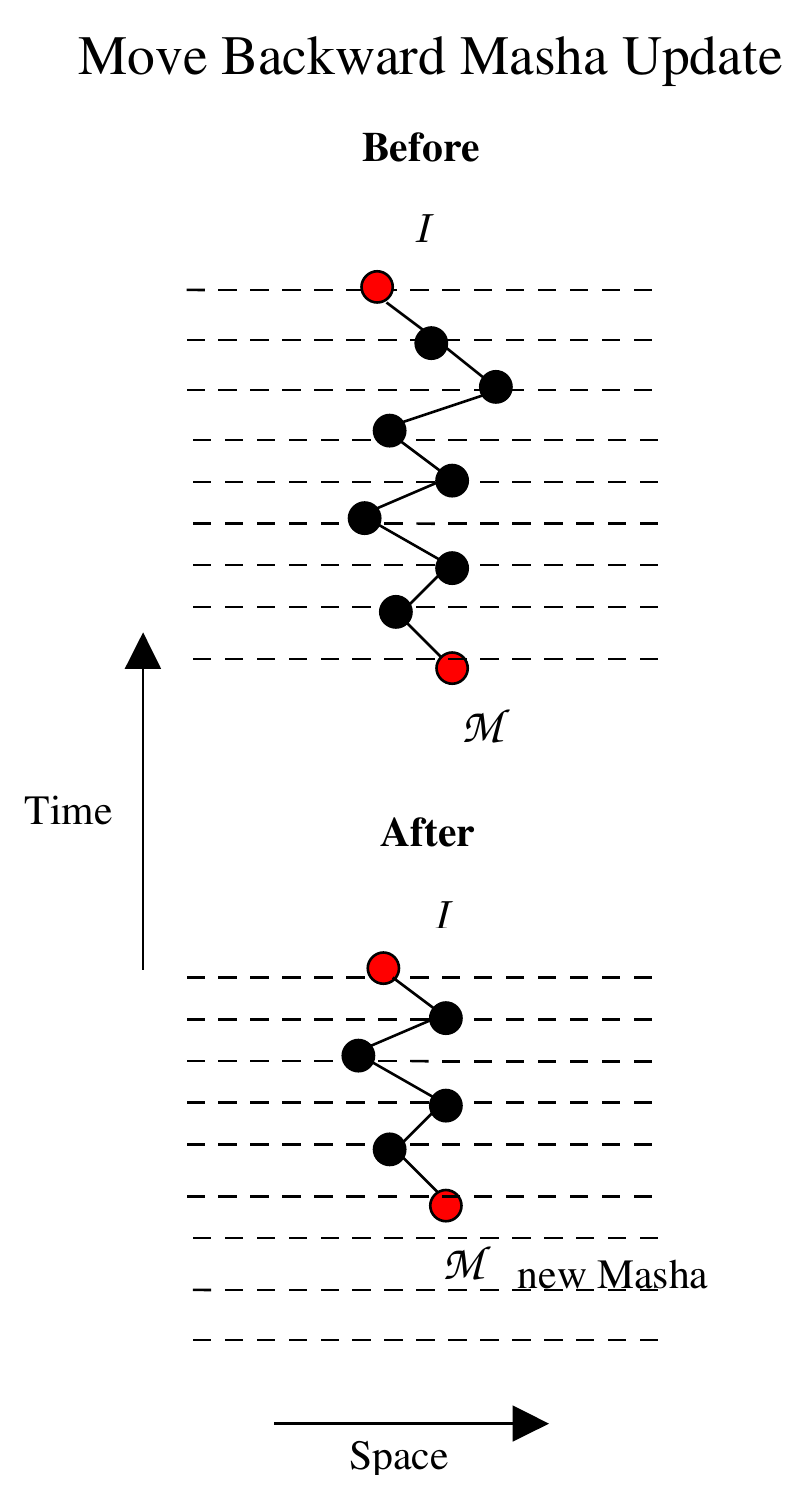}
\caption{\baselineskip 0.5cm As in Fig.\ref{fig:MoveForwardMasha}; 
but for moving \masha\ backward}\label{fig:MoveBackwardMasha}
\end{figure}

\hs Here \masha\ is moved forward in time as in 
Fig.\ref{fig:MoveBackwardMasha}. In other words, a chain of new
beads is erased forward in time beginning with the old {\it Masha}
until the erasure stops at a new worm-beginning which becomes
then the new {\it Masha}. The probability of this update is

\begin{equation}
P_{re, \cal M}\,=\,min\left\{1, e^{\Delta U-\mu M\epsilon}\right\},
\end{equation}

and is coded

\begin{equation}
P_{re, \cal M}\,=\,e^{-\mu\epsilon M+\Delta U} w_{lc,{\cal M}} 
\cdot w_t.
\end{equation}

As a safety measure, any bead which has a different type than
$\cal M$ causes the program to stop, as in (\ref{for90:stop.bead.M}).

\subsubsection{Move Backward Ira}

\begin{figure}[t!]
\includegraphics*[width=7cm,viewport=94 242 317 685]
{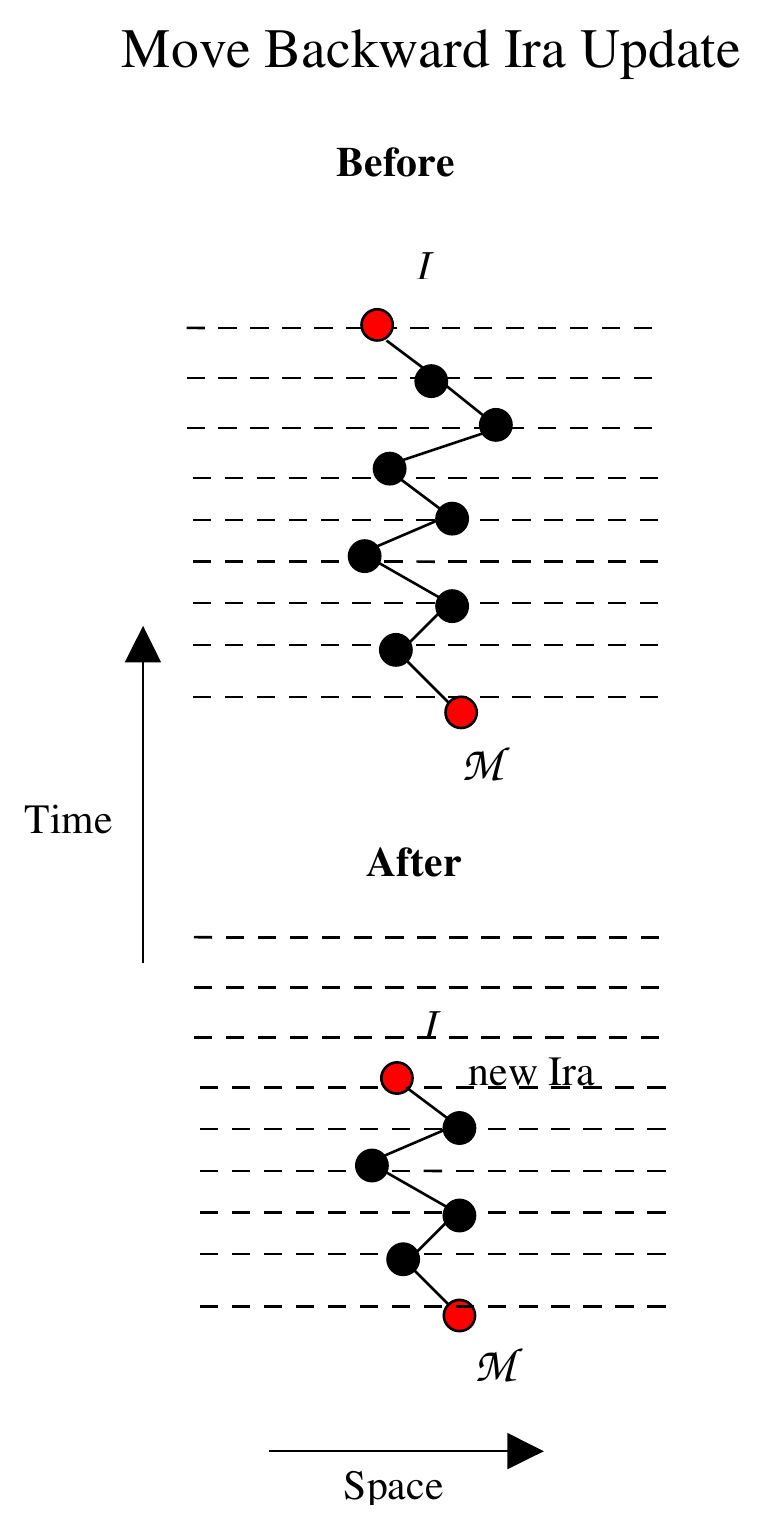}
\caption{\baselineskip 0.5cm As in Fig.\ref{fig:MoveForwardIra}; 
but for moving \ira\ backward.}\label{fig:MoveBackwardIra}
\end{figure}

\hs This update moves \ira\ backward in time as in 
Fig.\ref{fig:MoveBackwardIra}. Correspondingly, a chain of beads
is erased backwards in time beginning with the old {\cal Ira} until
the erasure stops at a new {\it Ira}. The resulting end of the worm
becomes the new {\it Ira}. The probability is given by

\begin{equation}
P_{re, \cal I}\,=\,min\left\{1,e^{\Delta U-M\epsilon \mu}\right\},
\end{equation}

and is coded:

\begin{equation}
P_{re, \cal I}\,=\,e^{-\mu\epsilon M+\Delta U}\cdot w_t \cdot 
w_{lc,\cal I}.
\end{equation}

Again, if a bead happens to have a different type than $\cal I$, 
the program terminates as in (\ref{for90:stop.bead.M})
as a safety measure.

\subsubsection{Glue}

\begin{figure}[t!]
\includegraphics*[width=6cm,viewport=93 294 292 702]{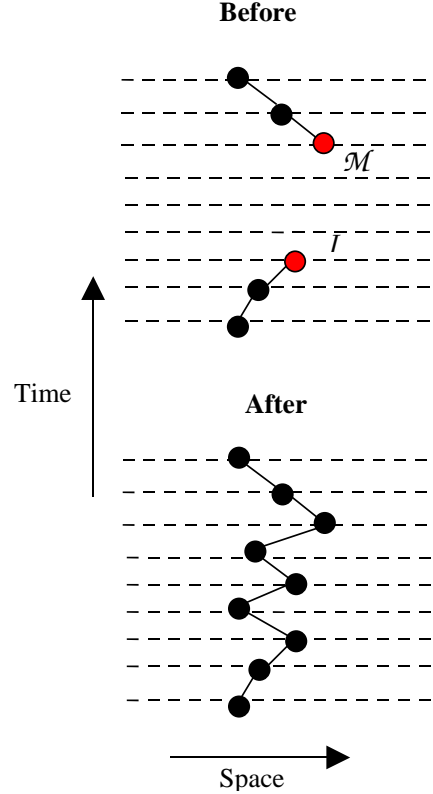}
\caption{\baselineskip 0.5cm Worm Algorithm update for closing a worm to become
a diagonal configuration.}\label{fig:glue}
\end{figure}

\hs Here, a worm of a type chosen in the INSERT update, is 
closed to become a ring polymer as shown in Fig.\ref{fig:glue}. 
{\it Masha} and {\it Ira} become ordinary beads in this case.
The probability for this update is

\begin{equation}
P_{glue}\,=\,min\left\{1,\frac{\rho_0(\mathbf{r}_{\cal I},
\mathbf{r}_{\cal M},M\epsilon)\,e^{\Delta U+\mu M \epsilon}}
{C \overline{M} N_{bd}}\right\}, \label{eq:Pglue}
\end{equation}

where $\mathbf{r}_{\cal I}$ and $\mathbf{r}_{\cal M}$ are the 
positions of $\cal I$ and $\cal M$, respectively, and $N_{bd}$ is 
the current total number of beads. The free-particle propagator $\rho_0$ 
is given by

\begin{equation}
\rho_0(\mathbf{r}_{\cal I}, \mathbf{r}_{\cal M}, M\epsilon)\,=\,
e^{-(\mathbf{r}_{\cal I}-\mathbf{r}_{\cal M})^2/(4 M\lambda \epsilon)}.
\end{equation}

Eq.(\ref{eq:Pglue}) is coded

\begin{eqnarray}
&&P_{glue}\,=\,\left[\frac{1}{4}\,e^{-\mu \epsilon M+\Delta U} 
e^{-(\mathbf{r}_{\cal I}-\mathbf{r}_{\cal M})^2\frac{mp}{2\epsilon M}} 
(aM)^{3/2} \right.\nonumber\\
&&\left.(N_{bd}+M-1) w_{ST} \overline{M}\frac{p_{gl}}{p_{cut}} \right. 
\left. w_{lc,{\cal I}} \cdot w_{lc,{\cal M}} \cdot w_t)\right]^{-1}, 
\nonumber \label{eq:Pglue-code}
\end{eqnarray}

where $a\,=\,2\pi\epsilon/mp$, $\mathbf{r}_{\cal I}$ and 
$\mathbf{r}_{\cal M}$ are the positions of {$\cal I$} and 
{$\cal M$}, $p_{gl}$ and $p_{cut}$ are the probabilities for attempting 
a glue or a cut, respectively. The cutting procedure is explained
in the next section below. Again, the glue beads must have the same 
type as the worm beads to be glued, otherwise the program stops using 
the Fortran statements similar to (\ref{for90:stop.bead.M}) or 
(\ref{for90:stop.bead.I}).

\subsubsection{Cut}\label{sec:cut}

\begin{figure}[t!]
\includegraphics*[width=6.5cm,viewport=89 282 301 684]
{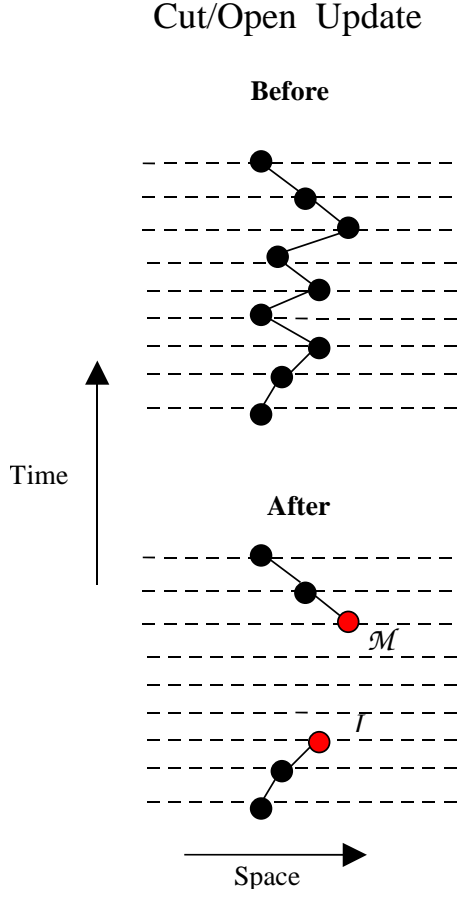}
\caption{\baselineskip 0.5cm As in Fig.\ref{fig:glue}; 
but for cutting a ring polymer open, i.e., making a 
diagonal configuration off-diagonal.}\label{fig:cut}
\end{figure}

\hs In this update, a randomly chosen piece of trajectory 
is removed from a ring polymer in order to create a worm as 
shown in Fig.\ref{fig:cut}. The beginning of the worm becomes
{\it Masha} and the end {\it Ira}. The mass is assigned according 
to the type of a randomly chosen bead ($nm$) using the code:

\begin{center}
\begin{eqnarray}
&&\mathbf{IF}\hspace{0.cm}(who_-are_-you(nm)\hspace{0.cm}\mathbf{.eq.}
\hspace{0.cm}\hbox{\bf .TRUE.})\hspace{0.cm}\mathbf{THEN} \nonumber\\
&&mp=m4\nonumber\\
&&\mathbf{ELSE\hspace{0.cm}IF}\hspace{0.cm}(who_-are_-you(nm)\hspace{0.2cm}
\mathbf{.eq.}\hspace{0.cm}\hbox{\bf .FALSE.})\hspace{0.cm}\mathbf{THEN}\nonumber\\
&&mp=m3\nonumber\\
&&\mathbf{ENDIF}
\end{eqnarray}
\end{center}

The probability for this update is given by

\begin{equation}
P_{cut}\,=\,min\left\{1,\frac{C\overline{M}N_{bd} 
e^{\Delta U-\mu M\epsilon}}
{\rho_0(\mathbf{r}_{\cal I},\mathbf{r}_{\cal M}, M\epsilon)}
\right\},
\end{equation}

and is coded

\begin{eqnarray}
&&P_{cut}\,=\frac{1}{4}\,e^{-\mu\epsilon M+\Delta U}
\,e^{(\mathbf{r}_I-\mathbf{r}_M)^2\cdot 
\frac{mp}{2\epsilon M}} (aM)^{3/2} \cdot\nonumber\\
&& N_{bd} \cdot w_{ST} \overline{M}\cdot \frac{p_{gl}}{p_{cut}} 
\cdot w_{lc,{\cal I}} \cdot w_{lc,{\cal M}} \cdot w_t
\end{eqnarray}

\begin{figure*}[t!]
\includegraphics*[width=12cm,viewport=118 365 520 704]
{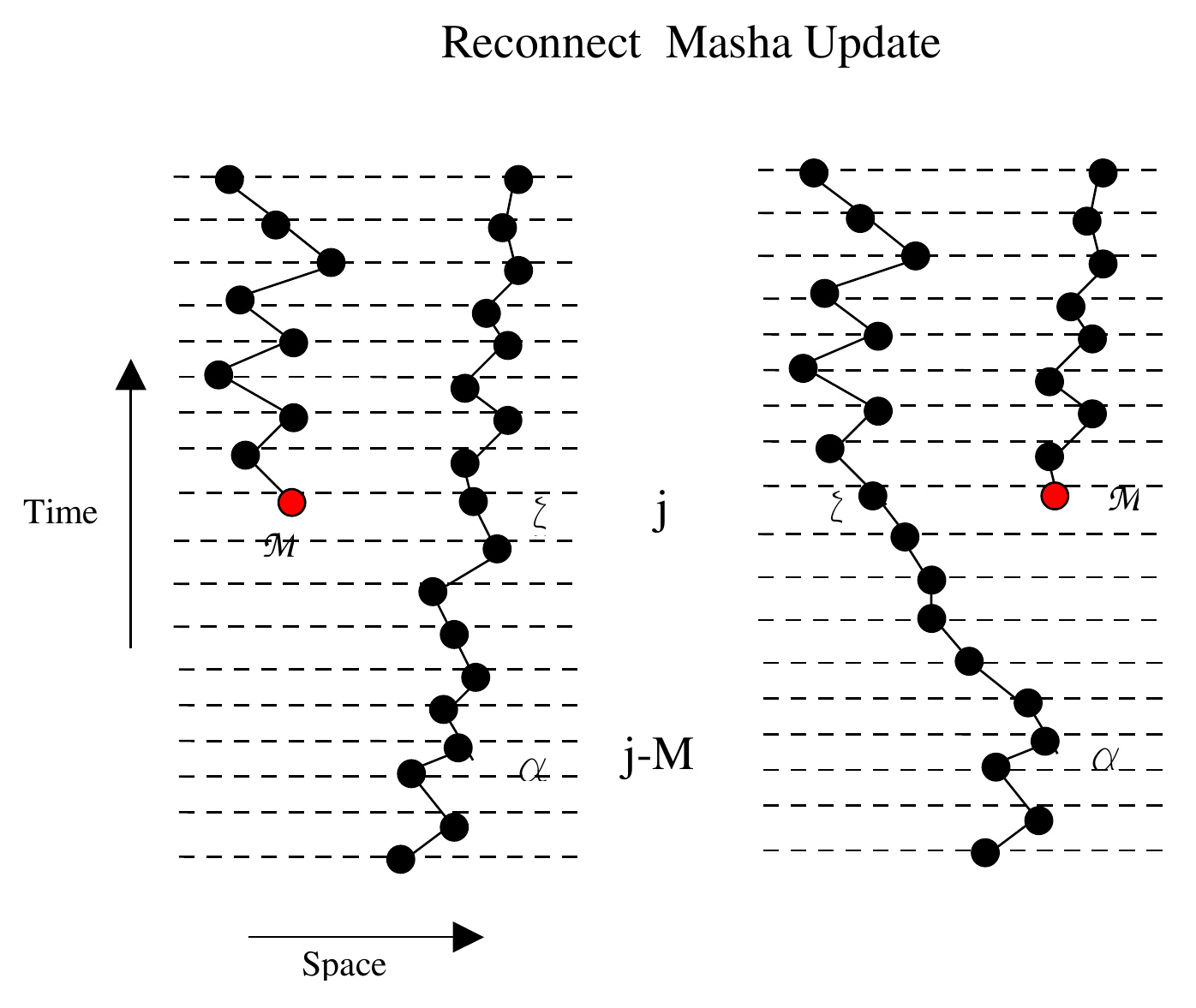}
\caption{\baselineskip 0.5cm Worm-Algorithm swap updates: 
After cutting a piece of trajectory between the beads $\alpha$ 
and $\xi$ from the path to the right of {$\cal M$}, \masha\ 
is reconnected to bead $\alpha$ chosen randomly on the other 
path, and $\xi$ becomes the new \masha.}
\label{fig:ReconnectMasha}
\end{figure*}

\subsubsection{Reconnect Masha}

\hs In this swap update, \masha\ of an open world line (worm) 
and time slice $j$ is connected to a randomly chosen bead $\alpha$ 
at time $j-M$ on another close world line (ring polymer) as shown in 
Fig.\ref{fig:ReconnectMasha}, by building a new trajectory between 
\masha\ at time $j$ and $\alpha$ at time $j-M$. Prior to this, 
the trajectory connecting $\alpha$ to a bead $\xi$, where $\xi$ 
is in the same time slice as \masha, is removed. Again, the mass 
of each bead is chosen depending on the type of worm inserted in 
Sec.\ref{sec:insert} and to be updated here. We made sure that the 
swap updates are done between the same type of beads as before:

\begin{eqnarray}
&&\mathbf{IF}\hspace{0.cm}(who_{-}are_{-}you({\cal M})\hspace{0.cm}
\mathbf{.ne.}\hspace{0.cm} who_{-}are_{-}you(\alpha))\hspace{0.cm}
\mathbf{RETURN}\nonumber\\
\end{eqnarray}

and throughout the removal of the trajectory (i.e., the beads say 
$\{bead1,\,bead2,\,bead3,\,\cdots,\,beadM\}$ between $\xi$ 
and $\alpha\equiv bead1$ one checks:

\begin{equation}
\mathbf{IF}\hspace{0.cm}(who_-are_-you(bead1)\hspace{0.cm}
\mathbf{.ne.}\hspace{0.cm}who_-are_-you(bead2))\hspace{0.cm}
\mathbf{STOP}
\end{equation}

and similarly for the rest of the beads, where 
$bead2=\mathbf{next}(bead1)$, $bead3=\mathbf{next}(bead2)$ and 
so on (see \cite{Boninsegni:06a}). Thus, if a bead does not have 
the same type as \masha\, the update is rejected. If the update 
is accepted, the previous $\xi$ then becomes the new \masha\ and
the old \masha\ is connected to $\alpha$. The probability for 
this update is

\begin{equation}
P_{re, \cal M}\,=\,min\left\{1, e^{-\Delta U}
\frac{\Sigma_{\cal M}}{\Sigma_\xi}\right\},
\end{equation}

where 

\begin{equation}
\Sigma_{\cal M}\,=\,\sum_{\sigma \epsilon {\cal L_M}} 
\rho_0(\mathbf{r}_{\cal M}, \mathbf{r}_{\sigma}, 
\overline{M}\epsilon),
\end{equation}

and

\begin{equation}
\Sigma_{\cal \xi}\,=\,\sum_{\sigma \epsilon {\cal L_M}} 
\rho_0(\mathbf{r}_{\cal \xi}, \mathbf{r}_{\sigma}, 
\overline{M}\epsilon),
\end{equation}

with $\cal L_M$ the list of particles in the slice $j-M$ in 
the bins that spatially coincide with the bin of \masha\, or 
one of its nearest neighbors, similarly for $\cal L_I$.

\hs The swap probability for \masha\ is coded

\begin{equation}
P_{re,\cal M}\,=\,\frac{\Sigma_{\cal M}}{\Sigma_{\cal \xi}}
\frac{w_{lc,\cal M}}{w_{lc,\cal \xi}} w_t e^{-\Delta U}
\end{equation}

with

\begin{equation}
\Sigma_{\cal M}\,=\,\left(\frac{1}{\sqrt{aM}}\right)^3\,
\sum_{i=1}^{h_m} e^{-(\mathbf{r}_{\cal M}-\mathbf{r}_i)^2 
\frac{mp}{2\epsilon M}},
\end{equation}

and

\begin{equation}
\Sigma_\xi\,=\,\left(\frac{1}{\sqrt{aM}}\right)^3\,
\sum_{i=1}^{h_m} e^{-(\mathbf{r}_\xi-\mathbf{r}_i)^2 
\frac{mp}{2\epsilon M}}\label{eq:sigmaxi}.
\end{equation}

Here $h_m$ is the number of particles in $\cal L_M$ and 
(in the next section) in $\cal L_I$.

\subsubsection{Reconnect Ira}

\hs This is a swap update as in the previous section but for \ira\ 
as shown in Fig.\ref{fig:ReconnectIra}. The probability for this 
update is given by

\begin{equation}
P_{re, {\cal I}}\,=\,min\left\{1, e^{\Delta U}\,\frac
{\Sigma_{\cal I}}{\Sigma_\xi}\right\},
\end{equation}

where 

\begin{equation}
\Sigma_{\cal I}\,=\,\sum_{\sigma \epsilon
{L}_J}\,\rho_\sigma(\mathbf{r}_I,\mathbf{r}_\sigma,\overline{M}
\epsilon),
\end{equation}

and $\Sigma_{\xi}$ was given by Eq.(\ref{eq:sigmaxi}) previously. 
The probability for this update is coded:

\begin{equation}
P_{re, {\cal I}}\,=\,\left(\frac{1}{\sqrt{a\overline{M}}}\right)^3\,
\frac{\Sigma_{\cal I}}{\Sigma_\xi} \frac{w_{lc,\cal I}}
{w_{lc,\cal \xi}}\,e^{\Delta U}\,w_t\exp(\Delta U).
\end{equation}

Again, one makes sure that the swap updates are done on the 
same type of beads:

\begin{figure*}[t!]
\includegraphics*[width=13cm,viewport=118 367 517 703]
{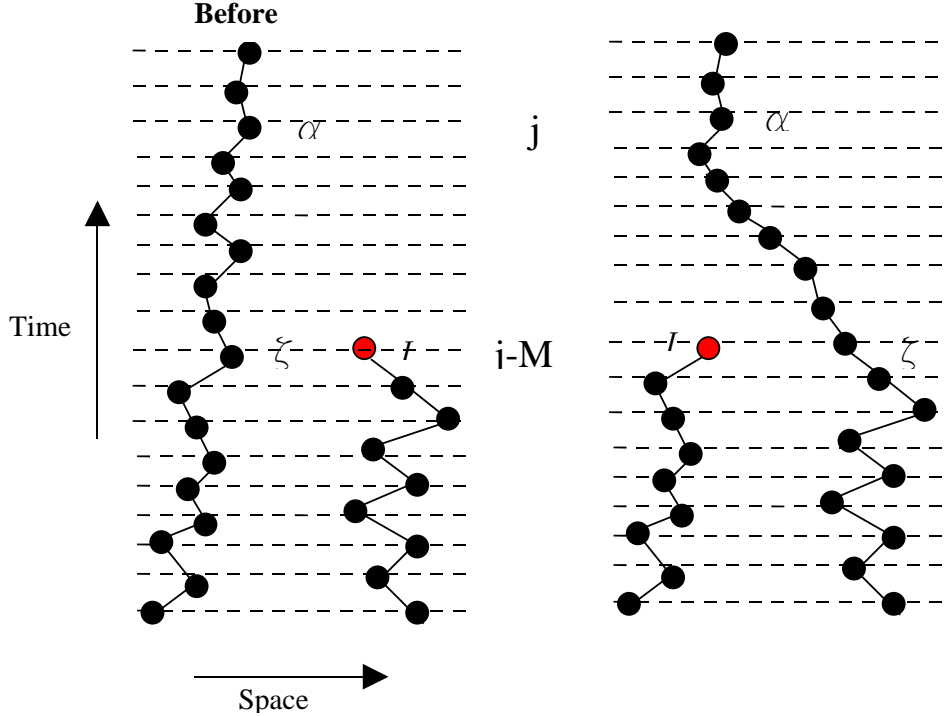}
\caption{\baselineskip 0.5cm Worm Algorithm swap update: 
\ira\ is reconnected to $\alpha$ after removing the trajectory 
between $\xi$ and $\alpha$. $\xi$ is at the same time as \masha.}
\label{fig:ReconnectIra}
\end{figure*}

\begin{equation}
\mathbf{IF}\hspace{0.cm}(who_-are_-you({\cal I})\hspace{0.cm}
\mathbf{.ne.}\hspace{0.cm}who_-are_-you(\alpha))\hspace{0.cm}
\mathbf{STOP},
\end{equation}

and during the removal of the path between $\alpha$ and $\xi$

\begin{equation}
\mathbf{IF}\hspace{0.cm}(who_-are_-you(bead1)\hspace{0.cm}
\mathbf{.ne.}\hspace{0.cm}who_-are_-you(bead2))\hspace{0.cm}
\mathbf{STOP}.
\end{equation}

\subsubsection{Insert Link}

\begin{figure}[t!]
\includegraphics*[width=6.5cm,bb=93 311 297 695]{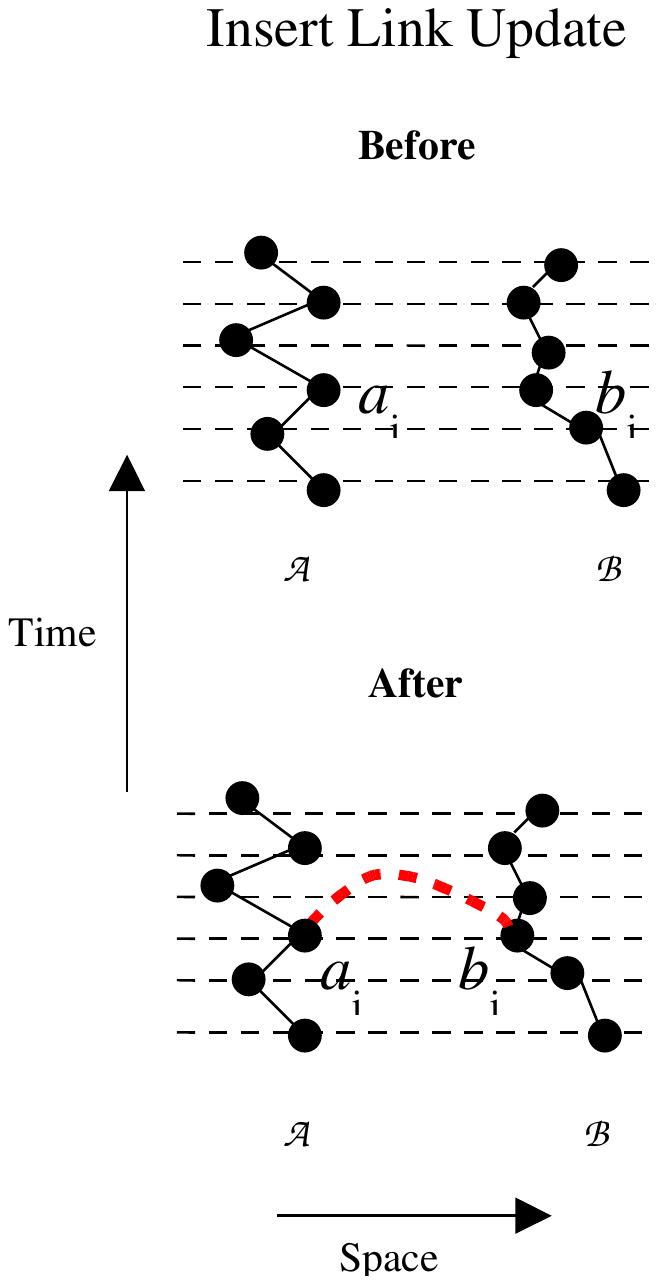}
\caption{\baselineskip 0.5cm Insert link update: A bead is created 
between randomly chosen beads $a_j$ and $b_j$ on different world 
lines and in the same time slice in their spacial bins $\cal A$ 
and $\cal B$, respectively.}\label{fig:InsertLink}
\end{figure}

\hs In addition to the previous updates, this update creates a bond 
(diagrammatic link) between the beads. In Fig.\ref{fig:InsertLink}, a 
bond (link) is created between beads $a_i$ and $b_i$ and the probability 
for this update is given by:

\begin{equation}
P_{crb}\,=\,\frac{(\overline{M}+1)\,n_B}{(\ell_{bnd}+1)\,
P_{\cal AB}}\,\left(e^{-fu(\mathbf{r}_{a_j}-\mathbf{r}_{b_j})}-1\right),
\end{equation}

where, $u(\mathbf{r}_{a_j}-\mathbf{r}_{b_j})$ is the interaction
potential between beads $a_j$ and $b_j$, $n_B$ is the number of 
beads in a spatial bin $\cal B$ within the slice $j$ of the bead 
$a_j$, where the update will be given a try, $\ell_{bnd}$ is the 
total number of bonds in the initial configuration, and $P_{\cal AB}$ 
is a probability that depends on the distance between bins $\cal B$ 
and $\cal A$. The probability is encoded

\begin{equation}
P_{crb}=pat\cdot (e^{-fu(\mathbf{r}_{a_j}-\mathbf{r}_{b_j})}-1)
\frac{\overline{M}+1}{n_{li}\,\cdot\,prob(evk)}
\end{equation}

where $n_{li}$ is the total number of links, $pat$ the number 
of beads $n_B$, $prob(evk)$ is $P_{\cal AB}$. Again, the type 
of bead to which a link is created doesn't matter. So, we do not
check here whether two beads to be linked have the same type or not.

\subsubsection{Remove Link}

\begin{figure}[t!]
\includegraphics*[width=6.5cm,bb=90 336 300 695]{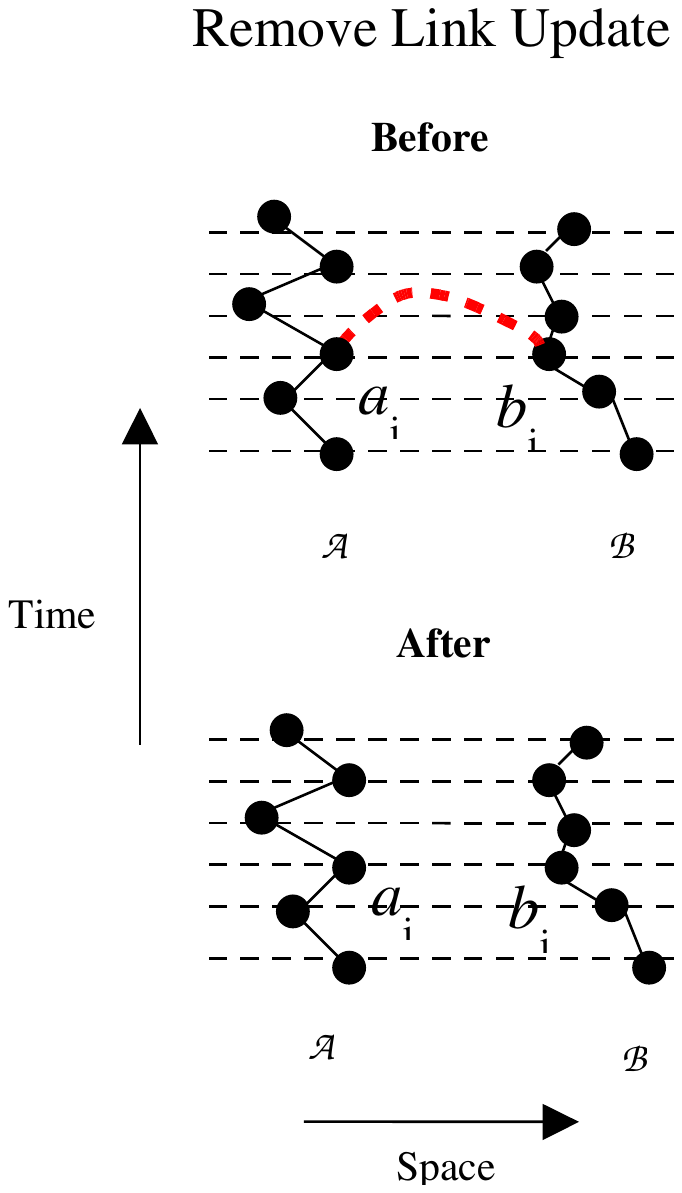}
\caption{\baselineskip 0.5cm As in Fig.\ref{fig:InsertLink}; but 
here a bond is removed.}\label{fig:Removelink}
\end{figure}

\hs This update removes a bond between beads $a_i$ and $b_i$, 
as shown in Fig.\ref{fig:Removelink}. The probability for this 
update is given by

\begin{equation}
P_{rmb}\,=\,\frac{\ell_{bnd} P_{\cal AB}} {(\overline{M}+1)\,n_B}\,
\left(e^{-fu(\mathbf{r}_{a_j}-\mathbf{r}_{b_j})}-1\right)^{-1}.
\end{equation}

\subsubsection{Diagonal}

\begin{figure*}[t!]
\includegraphics*[width=13cm,bb=119 316 515 698]{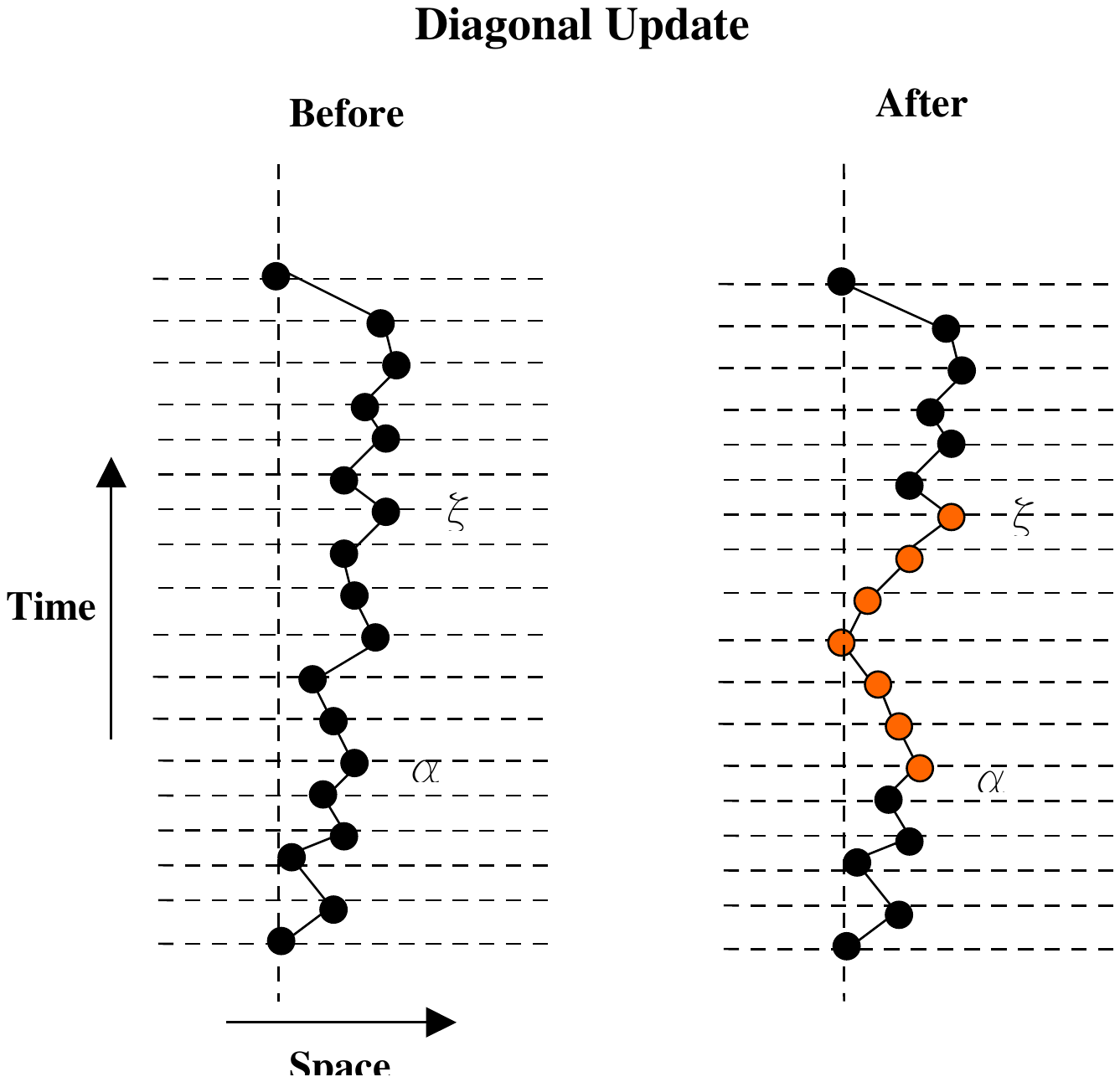}
\caption{\baselineskip 0.5cm Worm Algorithm Diagonal update where 
a piece of a trajectory is replace by another one, necessarily 
of the same type of beads.}\label{fig:diagonal}
\end{figure*}

\hs In this update, a randomly chosen piece of trajectory is 
removed from a closed path and replaced by a newly generated 
trajectory, as shown in Fig.\ref{fig:diagonal}. The probability 
for this update is given by

\begin{equation}
P_{diag}\,=\,e^{-\Delta U}.
\end{equation}

The newly generated trajectory must have the same type as the 
initial diagonal configuration, otherwise the update is rejected.

\subsection{Mobility of \xhe3\ in \zhe4}

\hs There is an inherent difficulty in the diffusion of \xhe3\
atoms in bulk \zhe4. To increase the mobility of \xhe3\ 
inside \zhe4, we applied an approach invented by previous 
authors \cite{Corboz:2008}, which makes use of the concept of a
fictitous or fake particle. On the other hand, this method also
addresses the diffusion of \zhe4\ atoms in the system. In this 
technique, one introduces into the system a fake \xhe3\ or \zhe4\ 
particle whose mass is allowed to vary during the simulation in 
increments of $\pm dm$. One can then increase the mobility of 
\xhe3\ and \zhe4\ atoms by reducing their mass or vice versa. 

\hs Computationally, an array $markf$ is introduced in order to 
mark beads as either fake ({\bf{.FALSE.}}) or real ({\bf{.TRUE.}}). 
This array is initialized in the beginning to {\bf{.TRUE.}}. 
Next, two mass differences

\begin{eqnarray}
&&\Delta m_3 = |m_{fake}-m_3| \nonumber\\
&&\Delta m_4 = |m_{fake}-m_4|
\end{eqnarray}

determine whether a fake \xhe3\ or \zhe4\ atom of mass $m_{fake}$ 
is to be chosen. The mass $m_{fake}$ is initialized to $m_3$ and 
then updated by a subroutine as explained below. If $|\Delta m_3|<dm$, 
where $dm=(m_4-m_3)/10.$, a subroutine choosing a fake \xhe3\
particle is called. Otherwise, if $|\Delta m_4|<dm$, another subroutine 
chooses a fake \zhe4\ particle (see Appendix). When a fake particle 
is chosen, the beads of its closed trajectory are labelled {\bf .FALSE.}. 

\subsubsection{Choosing a fake \xhe3\ particle}

\hs In the subroutine choosing a fake \xhe3\ particle, a bead 
($bead1$) is selected randomly from a list of beads [$nlist()$]:

\begin{equation}
i=rndm()*nmnm+1;\hspace{0.5cm} ip=nlist(i);\hspace{0.5cm}bead1=ip,
\label{eq:choose_bead_mobility} 
\end{equation}

where $nmnm$ is the number of beads at some number of Monte Carlo 
steps. If it happens that ($bead1$) is a \xhe3\ atom, a trajectory
of length $\beta=Mbeta$ time slices is assigned using a bead-list 
array $lbfnew()$ starting with $lbfnew(0)=bead1$. Otherwise, 
if $bead1$ is \zhe4, the routine returns to 
(\ref{eq:choose_bead_mobility}) above and tries again until a 
\xhe3\ $bead1$ is chosen. If the last bead ($ip=lbfnew(Mbeta)$) 
is not equal to $bead1$, that is the particle is in an exchange 
cycle, the chosen fake trajectory is rejected, i.e., its beads 
are not relabelled {\bf .FALSE.}. The subroutine then returns 
to Eq.(\ref{eq:choose_bead_mobility}) and starts all over again. 
If all goes well, that is by having a fake and closed pure \xhe3\ 
or \zhe4\ trajectory, a loop labels the beads of the chosen 
trajectory by {\bf.FALSE.} to make it fake: 

\begin{eqnarray}
&&markf=\mathbf{.TRUE.}\nonumber\\
&&do\hspace{5mm}k=0,Mbeta\nonumber\\
&&lbf(k)=lbfnew(k)\nonumber\\
&&markf(lbf(k))=\mathbf{.FALSE.}\nonumber\\
&&enddo
\end{eqnarray} 

The subroutine choosing a fake \zhe4\ particle is exactly 
the same, except for \zhe4. This subroutine is called when 
$\Delta m_4\le dm$, i.e., when $mfake$ has reached the mass 
of \zhe4\ during the mass update described next. Once a fake 
trajectory is chosen, its mass is updated by a subroutine 
for changing the mass of the fake particle. Physical properties
are then measured when $|\Delta m_3|\le dm$ or $|\Delta m_4|\le dm$.
Hence, any trajectory which has $\Delta m_3<dm$ or $\Delta m_4<dm$
is considered real and can be used to measure physical properties
in a given particle number sector. Thus when $\Delta m_3<dm$, 
the routine looks for another \xhe3\ atom to put the fake label
on, i.e., one looks for the bead which is the same as the current
fake, not in exchange cycles and not fake. Once this bead is
found, the previous fake labels are dropped and given to the new
bead upon which a whole new closed trajectory is labelled fake
to which this beads belongs. Similarly, when $\Delta m_4<dm$, the 
same procedure is applied, except that one chooses a fake \zhe4\ 
atom. A fake atom is not introduced when a worm is present. That 
is, one cannot perform these updates on worms, and one cannot have 
a fake worm. We must nevertheless emphasize that there will always 
be one fake atom in the configuration, it never disappears. And 
this fake atom is not part of any exchange cycle.

\subsubsection{Mass update}

\hs Once a fake trajectory has been selected, its mass is updated
using a subroutine (see Appendix) that we wrote for this 
following Ref.\cite{Corboz:2008}. In this subroutine, the trajectory 
mass is incremented or decremented in steps of $dm$, that is,

\begin{equation}
mfake=mold+sgn*dm, \label{eq:mass_update}
\end{equation}

where the sign of the increment, $sgn=\pm1$, is chosen randomly 
by the mechanism
\begin{eqnarray}
&&x=rndm()\nonumber\\
&&sgn=(-1)**(int(2.*x)),\nonumber\\
\end{eqnarray}

and $mold$ is the fake (old) mass from the previous update. Thus 
$mfake$ is constantly updated until it becomes either $m_4$ or 
$m_3$ within a small margin of error $|\Delta m_3|<dm$ or 
$|\Delta m_4|<dm$. In this case, the mass update stops momentarily 
allowing a measurement of physical properties. Then, a new fake 
trajectory is selected. We need to emphasize that the previous 
trajectory is reset to real ({\bf.TRUE.}) before either one of the 
subroutines for choosing a fake is called again. That is, no more than 
one fake trajectory is allowed. Further, inside the subroutine for 
choosing a fake mass, its mass is not allowed to obtain values less 
than $m_3$ or larger than $m_4$. If it reaches one of them, the mass 
update is rejected and $mfake$ is reset to $mold$. That is $mfake$ 
must always remain in the interval $[m_3,m_4]$. The mechanism by which 
the mass update in Eq.(\ref{eq:mass_update}) is accepted or rejected 
is according to a certain probability given by

\begin{equation}
P=\exp[\ell_k\Delta m/(2\epsilon)]
\cdot\exp[\alpha(m\pm\Delta m)]/\exp[\alpha m], 
\label{eq:Mass_prob}
\end{equation}

which is actually a modified version of that of Corboz \ea\ 
\cite{Corboz:2008} and which proved suitable for our purposes.
Here $\ell_k$ is defined as

\begin{equation}
\ell_k=\sum_{k=1}^M\,(\mathbf{r}_k-\mathbf{r}_{k-1})^2,
\end{equation}

and $\alpha$ is an adjustable parameter. According to this 
probability, if $P<1$ and $P>\xi$, where $\xi$
is a random number, the mass update is rejected and the newly
proposed fake mass in (\ref{eq:mass_update}) is set back to the
previous one, $m_{fake}=m_{old}$. Otherwise, $m_{fake}$ is assigned
the newly proposed value.

\subsubsection{Mass histogram}

\hs During the above processes, statistics for a mass histogram
for the several fake particles are collected in 10 mass bins as
was done in Ref.\cite{Corboz:2008}. This is in order to make sure
that the different 10 mass intervals are addressed with almost 
the same probability. For this purpose, one tunes the $\alpha$ 
value above such that one gets an almost mass flat histogram.

\section{Results and Discussion}\label{sec:resanddis}

\hs In this section, we present the results of our simulations.
We display the pair correlation function $g(r)$ for the three
different temperatures $T=30$, 40, and 50 mK, noting that the
correlations weaken as the temperature is reduced to 30 mK. Next,
the Matsubara Green's function \cite{Mahan:1990, Boninsegni:06a}
reveals the presence of a condensate fraction in the system,
whereas the \xhe3\ component completely depletes the superfluid.
In what follows, we first outline the difficulties which restricted
our investigations to only three temperatures.

\subsection{Difficulties in the WAQMC Simulations}

\hs It was possible to conduct WAQMC simulations on three 
milli-Kelvin temperatures only. The reasons are as follows. First,
in order to reach the milli Kelvin regime $T < 100$ mK, one needs
to use a large number of ``time" slices $\beta$ given by 
$M\,=\,\beta/\tau$. For our present purposes, we used a time 
step of $\tau=1/400\,\hbox{K}^{-1}$ and a simulation box of 
dimensions 19.693 $\AA\times$ 17.054 $\AA\times$ 26.798$\AA$.
For example, for $\beta\,=\,(1/0.04)$ K$^{-1}$ and 
$\tau\,=\,(1/400)$ K$^{-1}$, one needs $M\,=\,10000$. This is 
a very large number of time slices for WAQMC, let alone PIMC. 
Until now, and to the best of our knowledge, no one has ever 
conducted PIMC calculations below 250 mK because of the considerable 
computational cost involved. Nevertheless, we decided to take 
this step to explore the physics of the current system in this 
difficult regime.

\hs Second, because we used a repulsive statistical potential 
\cite{Pathria:1996} for the \xhe3\ pair interaction, the 
probabilities for worm updates on the \xhe3\ system were 
lowered substantially (as one can see by inspecting the 
worm-update probabilities in Sec.\ref{sec:worm-updates},
which are governed by the interaction of a worm with the rest of 
the system). Consider further the substantial large number of 
\xhe3\ atoms present in the current system which provides a large 
repulsive interaction energy. As a result, the evolution of the 
current simulated system took a considerable computational time 
in order to reach thermal equilibrium. The fact that the use of 
repulsive potentials in the WAQMC method can render the simulation 
inefficient was already mentioned by Boninsegni \ea\ 
\cite{Boninsegni:06b}. In other words, under these circumstances, 
the worm updates occur at a significantly lower rate.

\hs Third, the exact adjustment of the chemical potential $\mu$
posed another challenge. The average number of particles 
$\langle N\rangle$ is allowed to vary by running the WAQMC simulation
in the grand canonical ensemble. When the system eventually thermalizes,
the number of particles, as determined by $\mu$, stabilizes after
a long run or thermal evolution time. It is very difficult to 
predict the number of particles to which the system would eventually
thermalize by guessing $\mu$ from the outset, i.e., the beginning
of a simulation. One can only conduct several runs at different
$\mu$ and the same $T$ in order to obtain various numbers of particles
corresponding to the chemical potentials used. Then, one can construct a 
``calibration curve" of $\langle N\rangle$ vs. $\mu$ for each $T$
within an acceptable error range of $\langle N\rangle$. That way
$\mu$ can be predicted $-$numerically speaking $-$ more reliably
for other nearby temperatures. Yet, this procedure is very 
time-consuming, given that one needs to wait for the system to
thermalize for each value of $\mu$ chosen. It could take months
to determine the correct $\mu$ with the computational resources 
that we have currently available. As a result, we chose to conduct
a {\it qualitative} investigation of this system by running the
WAQMC simulations in the {\it canonical ensemble} by choosing
a reasonable $\mu$. In fact, it was later found that in the
milli-Kelvin temperature regime, $N$ turns out to be independent of
$\mu$.

\begin{figure}[t!]
\includegraphics*[width=8.5cm,viewport = 160 499 502 773,clip]
{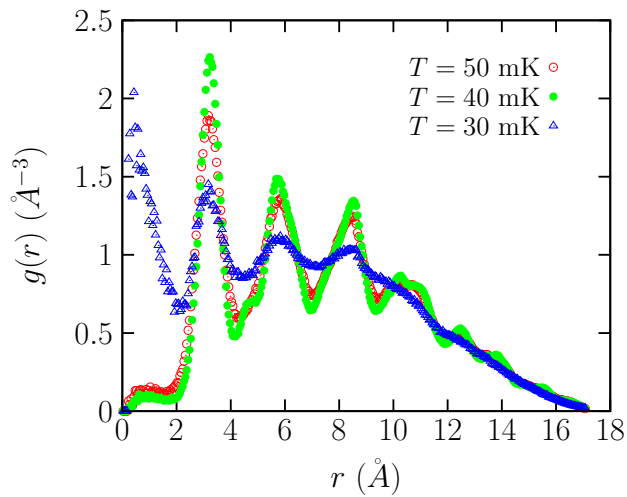}
\caption{WAQMC pair correlation function $g(r)$ for the \xsand\ system
at three different temperatures: $T=30$ mK (open circles); 
40 mK (solid circles); and 30 mK (open triangles). Distances are in
units of $\AA$, and $g(r)$ is in units of $\AA^{-3}$.}
\label{fig:plot.corr.vsT}
\end{figure}

\subsection{Pair correlations}

\hs The correlation function $g(r)$ counts the number of atom pairs
with interparticle distance $r$. It provides evidence for the 
clusterization of particles around certain locations in the system.
Fig.\ft\ref{fig:plot.corr.vsT} displays correlation functions for our
system at the indicated temperatures: 50 mK (open circles); 40 mK 
(solid circles); 30 mK (open triangles). The peaks in this figure
strongly indicate the presence of clusters $-$possibly droplets. This
explanation is similar to that given by Boninsegni and Szybisz 
\cite{Boninsegni:2004}, who investigated helium films on lithium
substrates at $T=0.5$ K. Their $g(r)$ acquires a nonzero value at
the origin, indicating that the helium film is forming droplets on 
the substrate surface. Inspecting Fig.\ft\ref{fig:plot.corr.vsT},
one can see that $g(r=0)=0$ at all $T$. That is, the \zhe4\ adsorbed
on the substrate forms no droplets, as it is almost a solid. The
rest of the peaks in $g(r)$ possibly signals the presence of pure
\zhe4\ clusters at $r\sim 3\AA$, \sandw\ (pair) clusters at 
$r\sim 6\AA$, and pure $^3$He-$^3$He (pair) clusters at $r\sim 8\AA$. 
This is a reflection of the zero-point motion of \xhe3\ and \zhe4, 
that of \xhe3\ being larger, of course. Accordingly, the pure \zhe4\ cluster
would have the lowest interparticle distances around $r\sim 3\AA$.
The \sandw\ cluster would have larger interparticle distances
because of the larger $^3$He zero-point motion. Finally,  the $^3$He
cluster has the largest interparticle distances as it is undergoing 
only \xhe3\ zero-point motion. Yet $g(r)$ in 
Fig.\ft\ref{fig:plot.corr.vsT} decays to zero at large $r\ge 16\AA$,
the reason being that our system is simulated in a box of finite 
size and does not extend to infinity. There are some remaining
oscillations in $g(r)$ at $r\ge 10\AA$, which could be indicative of
other types of structures. However, at $T=30$ mK, $g(r)$ has a peak
at $r\sim 0.5\AA$. Some particles may have left the higher layers
and approached the graphite surface, most likely \xhe3. Being 
attracted by the strong graphite potential, once the \xhe3\ atoms
reach the surface of the substrate, the strong \xhe3-graphite 
interaction ($\sim\,-200$ K) overcomes their zero-point motion 
($\sim 7$ K), and they begin to form more \xhe3\ or \sandw\ clusters
close to the surface. Further, the intensity of $g(r)$ at $r\sim 3$,
6, and $8\AA$ indicates clustering closer to the graphite surface,
as atoms leave the higher layers and approach the substrate.

\hs A question arises as to the role of temperature reduction on 
particle promotion and demotion from one layer to another. Are 
\xhe3\ atoms (or \zhe4) being demoted from the highest layer down, 
closer to the graphite surface? What is the role of the statistical 
potential in this case? We know that it is temperature-dependent.

\subsection{Matsubara Green's function}

\hs In what follows, we explore the possibility for the presence
of excitations in the system by measuring the Matsubara Green's 
function (MGF) $G(p,\tau)$ \cite{Mahan:1990} at zero momentum 
using WAQMC. In other words, we check whether our system, as 
simulated by WAQMC, has really reached its ground state or not. 
This is a crucial point in the verification of the reliability of 
the results. Often, in heavy computational techniques like WAQMC, 
such a step can give the green light for finally stopping the simulation. 

\begin{figure}[t!]
\includegraphics*[width=8.5cm,viewport = 190 523 554 773,clip]
{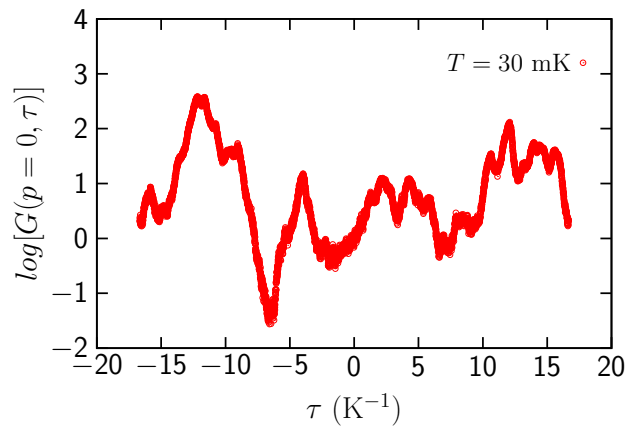}
\caption{The logarithm of the WAQMC zero-momentum Matsubara Green function 
$\log[G(p=0,\tau)]$ at $T=30$ mK. The ``time" $\tau$ is in units of K$^{-1}$.}
\label{fig:plotGGT0d03K}
\end{figure}

\begin{figure}[t!]
\includegraphics*[width=8.5cm,viewport = 184 523 554 773,clip]{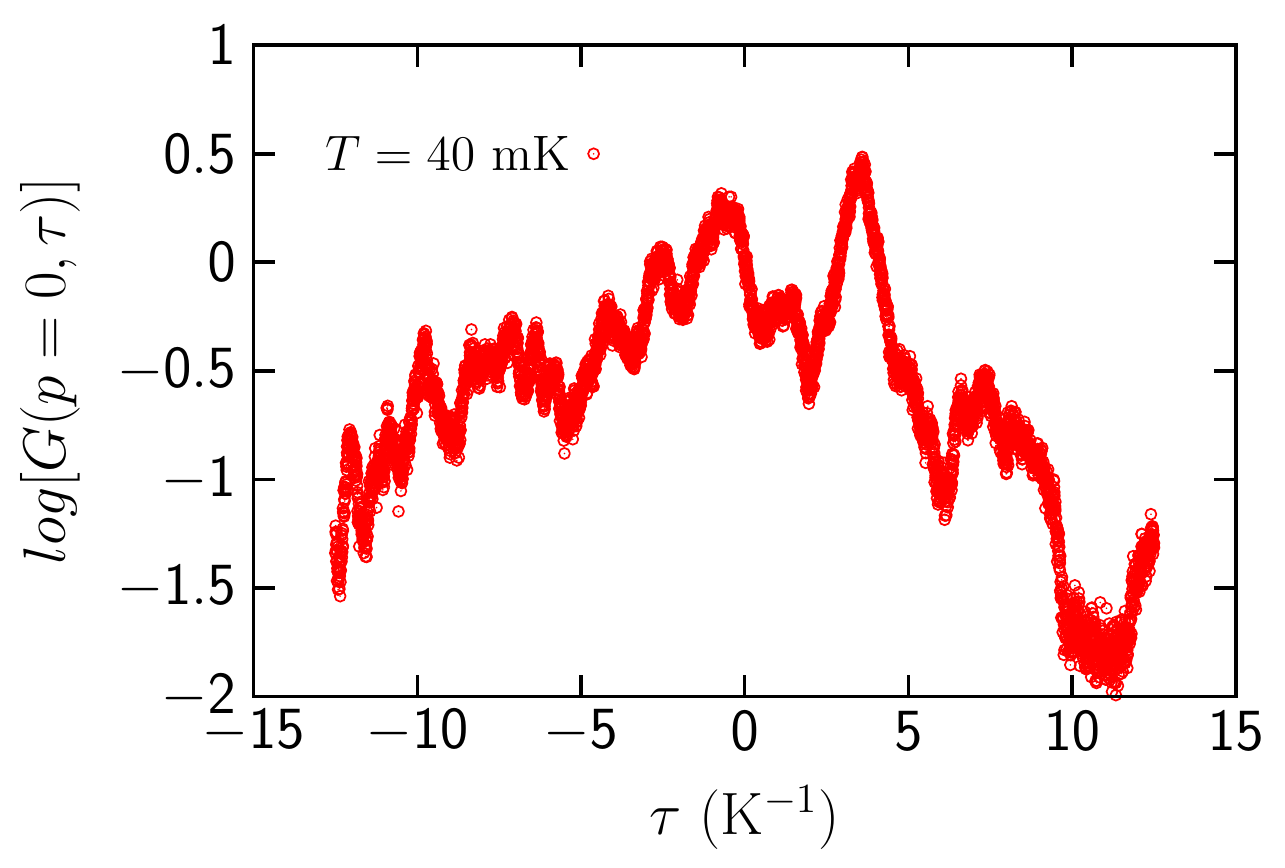}
\caption{As in Fig.\ft\ref{fig:plotGGT0d03K}; but for $T=40$ mK.}
\label{fig:plotGGT0d04K}
\end{figure}

\begin{figure}[t!]
\includegraphics*[width=8.5cm,viewport = 183 523 554 773,clip]{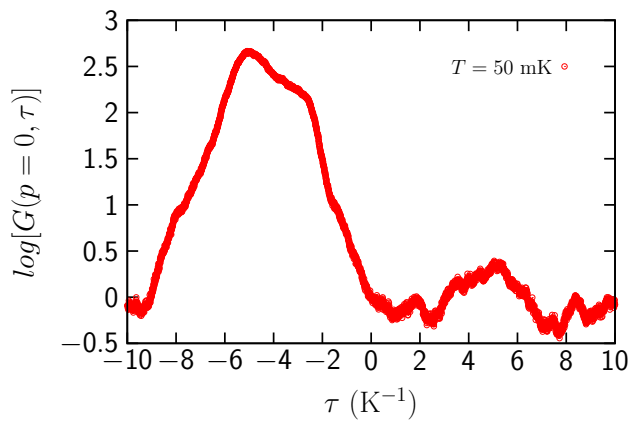}
\caption{As in Fig.\ft\ref{fig:plotGGT0d03K}; but for $T=50$ mK.}
\label{fig:plotGGT0d05K}
\end{figure}

\hs Figs.\ft\ref{fig:plotGGT0d03K}, \ref{fig:plotGGT0d04K}, and
\ref{fig:plotGGT0d05K} present the WAQMC $G(p=0,\tau)$ at $T=30$,
40, and 50 mK in the ``time" range $-\beta\le\tau\le\beta$.
The $G(p=0,\tau)$ signal significant activity in the system at
the various times $\tau$. The particles seem to propagate at
various amplitudes of the MGF in the $p=0$ state at
the different values of $\tau$; yet no signals for particle
excitations or deexcitations are detected. In fact, the Green
function at $\tau=0$ corresponds to the number of particles in
the condensate $N_0$! That is, according to Mahan \cite{Mahan:1990},
$G(p=0,\tau=0)\propto -N_0$, where the proportionality sign
arises because the Green function obtained in this treatment 
contains signals from both the fermions and the bosons. Accordingly,
one might be tempted to argue that there is a condensate in our 
system since, at $\tau=0$, the Green function in all three Figs.
\ft\ref{fig:plotGGT0d03K}, \ref{fig:plotGGT0d04K}, and
\ref{fig:plotGGT0d05K} displays a nonzero value. 

\subsection{Structure Factor}

\begin{figure}[t!]
\includegraphics*[width=8.5cm,viewport=153 486 532 775]{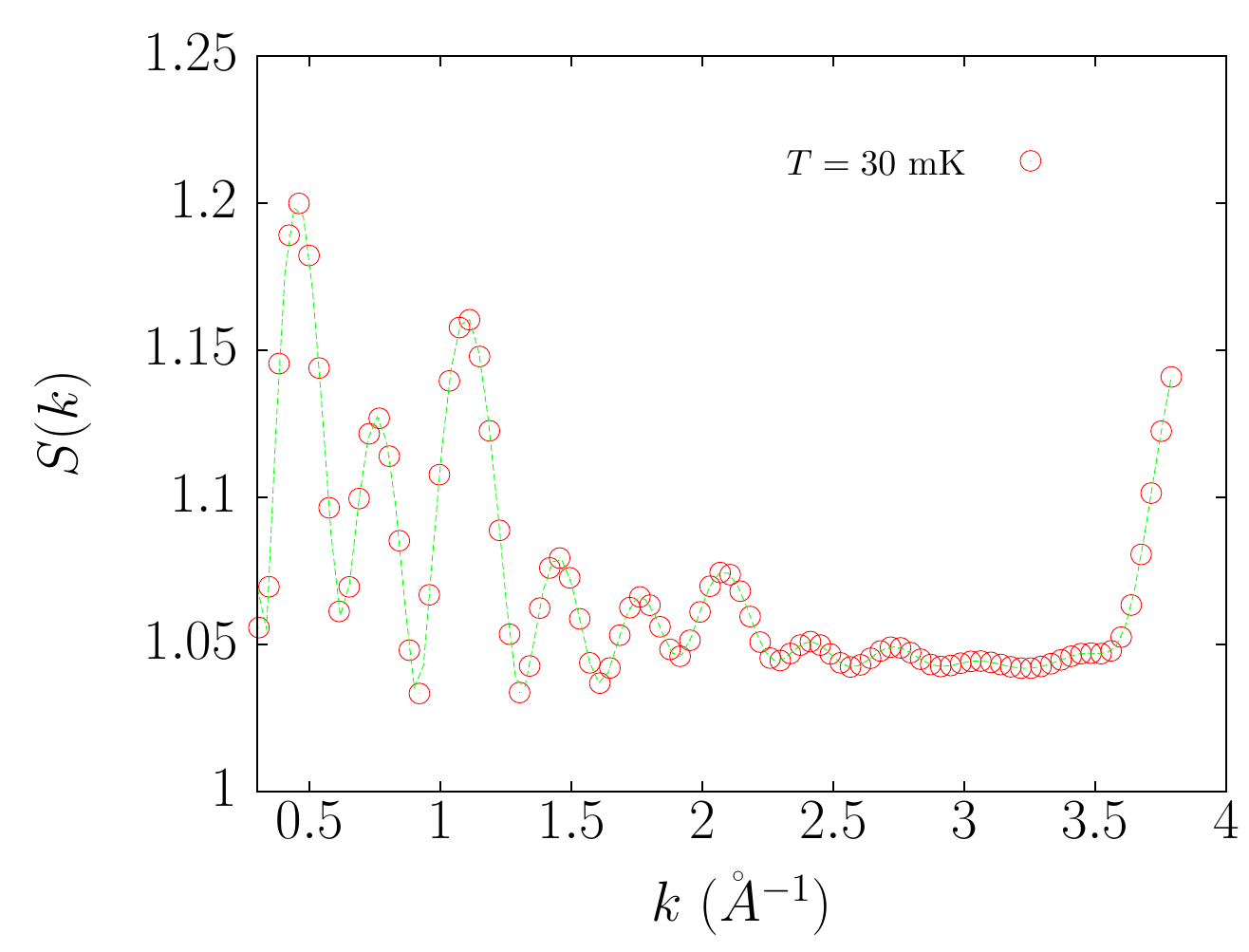}
\caption{WAQMC static structure factor $S(k)$ at $T=30$ mK. The quasi-momentum
wave vector $k$ is in units of $\AA^{-1}$.}
\label{fig:plot.SQ.T30mK}
\end{figure}

\hs Fig.\ft\ref{fig:plot.SQ.T30mK} displays the static structure 
factor $S(k)$ 
for the sandwich system at $T=30$ mK. Three significant Bragg peaks appear at 
$k\sim 0.5$, 0.75, and 1.2$\AA^{-1}$, which reveal crystalline order 
in the system, largely present in the first few \zhe4\ layers closest 
to the graphite substrate. The strong attraction of the helium atoms 
to the graphite forces crystalline order as the \zhe4\ atoms get adsorbed 
on the substrate surface. The absence of Bragg peaks in the higher layers 
is a consequence of the He-graphite potential becoming weaker. As a
result, the bulk \xhe3\ component is completely disordered.

\subsection{Density Profiles}

\begin{figure}[t!]
\includegraphics*[width=8.5cm,viewport=88 503 528 741]{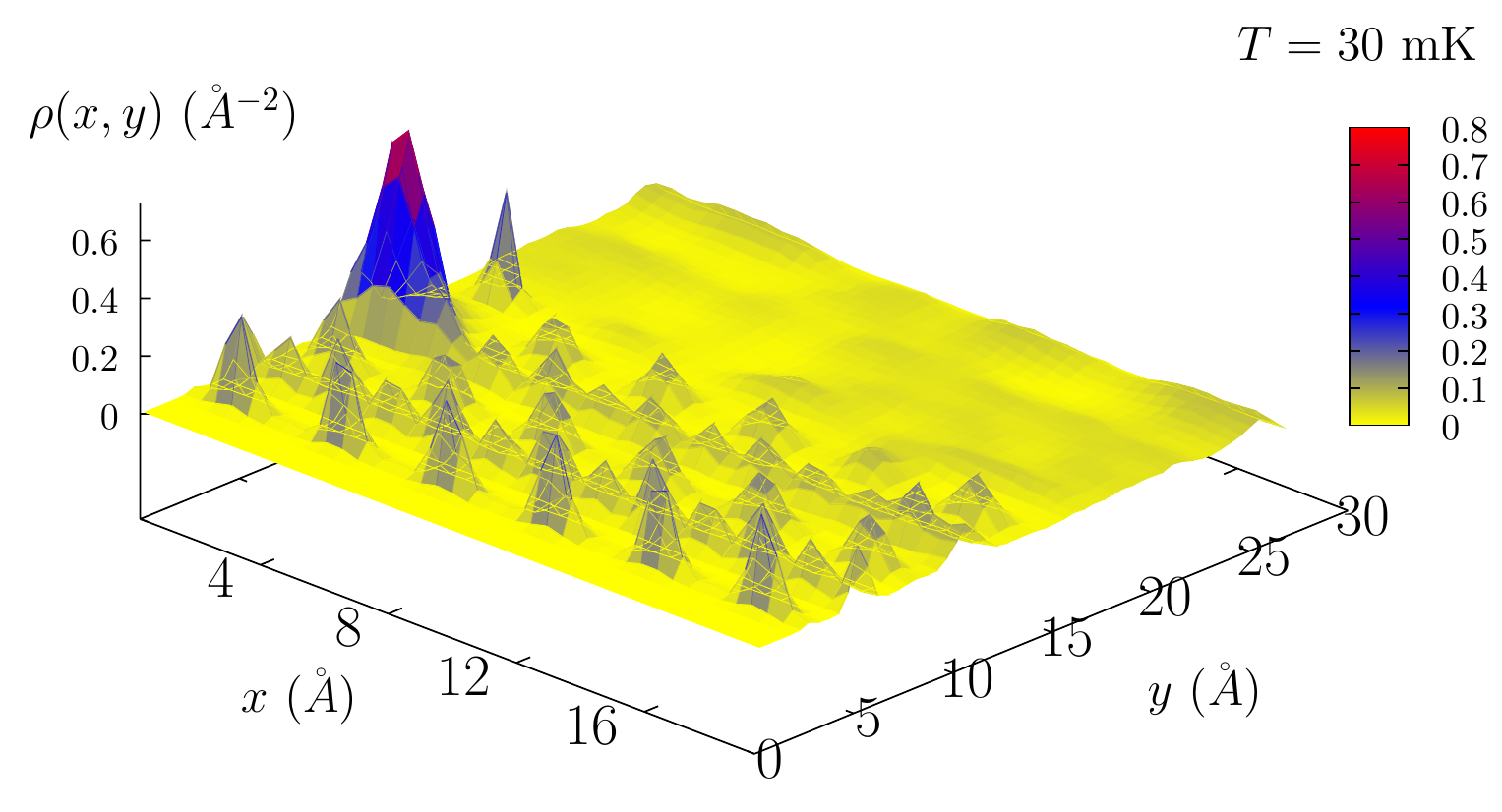}
\caption{Integrated density $\rho(x,y)$ at $T=$ 30 mK along the 
$z-$axis in the $xy$ plane perpendicular to the graphite substrate plane}
\label{fig:plot.DDmap.T30mK}
\end{figure}

\begin{figure}[t!]
\includegraphics*[width=8.5cm,viewport=88 503 528 741]{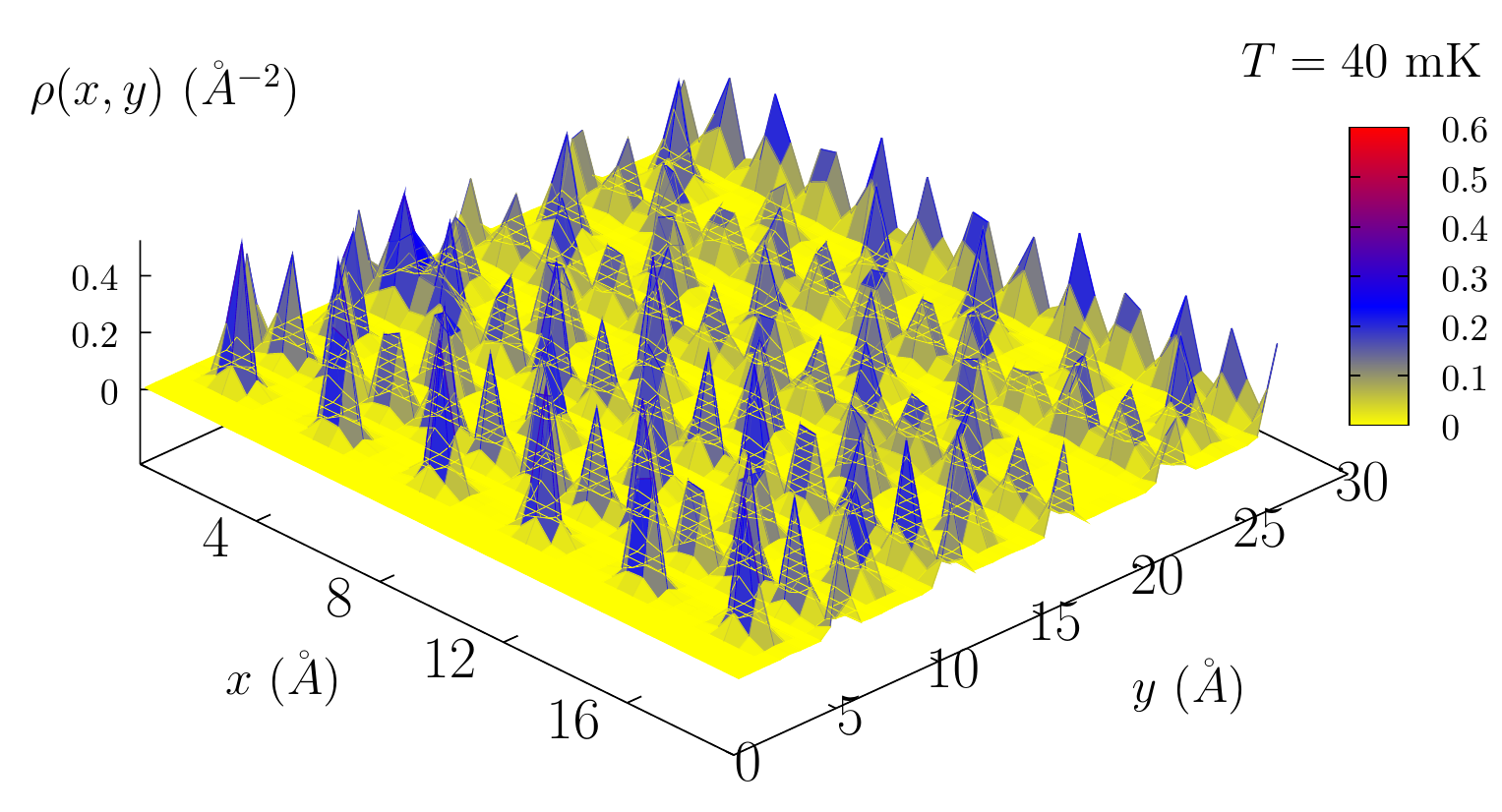}
\caption{As in Fig.\ft\ref{fig:plot.DDmap.T30mK}; but for 40 mK.}
\label{fig:plot.DDmap.T40mK}
\end{figure}

\begin{figure}[t!]
\includegraphics*[width=8.5cm,viewport=88 503 528 741]{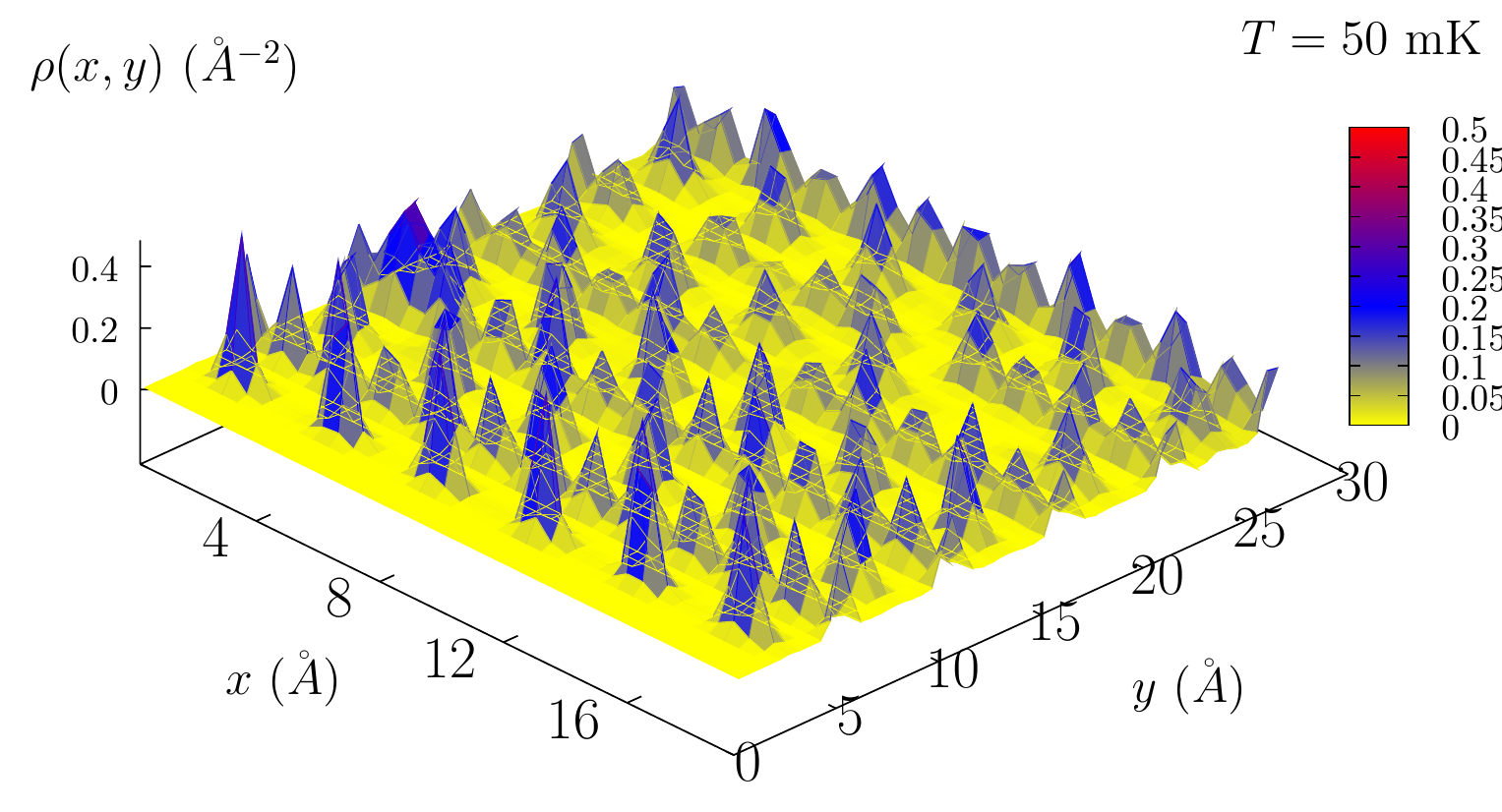}
\caption{As in Fig.\ft\ref{fig:plot.DDmap.T30mK}; but for 50 mK.}
\label{fig:plot.DDmap.T50mK}
\end{figure}

\hs Figures \ft\ref{fig:plot.DDmap.T30mK}-\ref{fig:plot.DDmap.T50mK} 
display integrated two-dimensional profiles at $T=30$, 40, and 50 mK, 
respectively, in the $x-y$ plane perpendicular to the graphite surface 
$x-z$. The integration is performed along the $z-$axis. A peculiar density 
distribution is observed at 30 mK, where there is a high peak observed 
(red cusp), indicating clusterization of the helium atoms. However, 
it is difficult to tell whether these would be \xhe3\ or \zhe4\ (or both) 
clusters. Further, there is a smooth, slightly wavy area in the xy plane 
at $20\le y\le 30\AA$ where a crystal structure seems to be absent, 
and may possibly indicate the presence of a liquid. 
Figure \ref{fig:plot.DDmap.T40mK}, on the other hand, does not reveal any 
signals for clusterization at 40 mK. The sharp, periodically ordered peaks 
are indicative of a largely prevalent crystalline structure. Figure 
\ref{fig:plot.DDmap.T50mK} reveals the same absence of crystallization.

\section{Conclusions}\label{sec:conclusions}

\hs In summary, then, the thermal and structural properties of the \sandw\
system were investigated at low temperatures in the milli-Kelvin regime.
These temperatures lie in an extremely difficult regime in which WAQMC
runs must take a long time so as to give good results. The correlations,
structure factor, Matsubara Green's function, and density profiles
were explored. A major point in this study is that we used a repulsive 
statistical potential in order to describe the \xhe3\ atoms as real 
fermions. Although this potential slowed down the evolution of the system 
during the WAQMC calculation $-$that is, the acceptance probability of 
worm-updates was reduced and occured less frequently than when using 
attractive interactions for the \xhe3\ atoms $-$ we were still able to 
evaluate the properties of the system.

\hs It was found that the superfluid fraction of the sandwich has zero
value. This is because the large number of \xhe3\ atoms depletes the
superfluid strongly. The correlation function of the system was evaluated
at different temperatures. It was found to display three peaks at 
$r\sim 3$, 6, and 8$\AA$, signalling ${}^4$He$-{}^4$He, $^4$He$-{}^3$He 
and ${}^3$He$-{}^3$He clusterizations, respectively. The structure factor
was then investigated at $T=30$ mK. It shows a quasicrystalline structure
up to $k\sim 2.5\AA^{-1}$; but then disorder sets in. Three significant
Bragg peaks appear at $k=0.5$, 0.75, and 1.2$\AA^{-1}$. The density 
profile of the system was explored at different temperatures. It was
shown to depend strongly on temperature. Furthermore, at $T=30$ mK, there
is a clustering of the \xhe3\ atoms in some region indicated by the
highest peak in Fig.\ft\ref{fig:plot.DDmap.T30mK} In the future, we will 
explore a few \xhe3\ atoms placed on a layer of \zhe4\ atoms adsorbed on 
graphite using the same WAQMC code modified here.

\acknowledgements

\hs We are very indebted for Nikolay Prokofev for providing
us with his Worm Algorithm code. We would also like to thank him 
for his help in the modification of the code for the present purpose,
and for enlighting and stimulating discussions. One of the authors
(HBG) is grateful for The University of Jordan for granting him a
sabbatical leave during which this work was completed. This research 
has been generously supported by the University of Jordan under 
project number 74/2008-2009 dated 19/8/2009..

\bibliography{wa}
\end{document}